\documentclass[12pt,preprint]{aastex}

 \newcommand{\feh}{\mathrm{[Fe/H]}}

\newcommand{\avevR}{\langle v_R \rangle}
\newcommand{\avevphi}{\langle v_\phi \rangle}
\newcommand{\avevz}{\langle v_z \rangle}
\newcommand{\kms} {\mathrm{km \ s^{-1}}}

\shorttitle{Local Halo Star Streams}
\shortauthors{Kepley et al.}

\begin{document}

  \title{Halo Star Streams in the Solar Neighborhood}
  \author{Amanda A. Kepley}
  \affil{Department of Astronomy, Case Western Reserve University,
  Cleveland, OH 44106, U.S.A.; and Department of Astronomy, University
  of Wisconsin--Madison, 475 North Charter Street, Madison, WI 53706, U.S.A.
\email{kepley@astro.wisc.edu}}
\author{Heather L. Morrison}
  \affil{Department of Astronomy, Case Western Reserve University,
  Cleveland, OH 44106, U.S.A. \email{ heather@vegemite.case.edu}}
  \author{Amina Helmi}
  \affil{Kapteyn Astronomical Institute, University of Groningen, P.O. Box 800, 9700 AV Groningen, Netherlands \email{ahelmi@astro.rug.nl}}
\author{T.D. Kinman}
\affil{National Optical Astronomy Observatory, P.O. Box 26732, Tucson
  AZ 85726, U.S.A.  \email{tkinman@noao.edu} }
  \author{Jeffrey Van Duyne}
  \affil{Department of Astronomy, Yale University, P.O. Box 208121,
  New Haven, CT 06520, U.S.A. \email{vanduyne@astro.yale.edu} }
  \author{John C. Martin} 
  \affil{University of Illinois at Springfield, 1 University Plaza, MS
  HSB 314, Springfield, IL 62703, U.S.A. \email{jmart5@uis.edu}}
  \author{Paul Harding}
  \affil{Department of Astronomy, Case Western Reserve University,
  Cleveland, OH 44106, U.S.A. \email{harding@dropbear.case.edu}}
\author{John E. Norris and Kenneth C. Freeman}
\affil{Research School of Astronomy and Astrophysics, Australian
  National University, Private Bag,
Weston Creek PO, 2611 Canberra, ACT, Australia \email{jen@mso.anu.edu.au,kcf@mso.anu.edu.au}}

\begin{abstract}

We have assembled a sample of halo stars in the solar neighborhood to
look for halo substructure in velocity and angular momentum space. Our
sample (231 stars) includes red giants, RR Lyrae variable stars, and
red horizontal branch stars within 2.5 kpc of the Sun with [Fe/H] less
than --1.0. It was chosen to include stars with accurate distances,
space velocities, and metallicities as well as well-quantified
errors. With our data set, we confirm the existence of the streams
found by Helmi and coworkers, which we refer to as the H99
streams. These streams have a double-peaked velocity distribution in
the $z$ direction (out of the Galactic plane). We use the results of
modeling of the H99 streams by Helmi and collaborators to test how one
might use $v_z$ velocity information and radial velocity information
to detect kinematic substructure in the halo. We find that detecting
the H99 streams with radial velocities alone would require a large
sample (e.g., approximately 150 stars within 2 kpc of the Sun and
within 20$^\circ$ of the Galactic poles). In addition, we use the
velocity distribution of the H99 streams to estimate their age. From
our model of the progenitor of the H99 streams, we determine that it
was accreted between 6 and 9~Gyr ago. The H99 streams have
[$\alpha$/Fe] abundances similar to other halo stars in the solar
neighborhood, suggesting that the gas that formed these stars were
enriched mostly by Type II supernovae. We have also discovered in
angular momentum space two other possible substructures, which we
refer to as the retrograde and prograde outliers. The retrograde
outliers are likely to be halo substructure, but the prograde outliers
are most likely part of the smooth halo. The retrograde outliers have
significant structure in the $v_\phi$ direction and show a range of
[$\alpha$/Fe], with two having low [$\alpha$/Fe] for their [Fe/H].
The fraction of substructure stars in our sample is between 5\% and
7\%. The methods presented in this paper can be used to exploit the
kinematic information present in future large databases like RAVE,
SDSSII/SEGUE, and {\em Gaia}.

\end{abstract}


\section{Introduction}

The kinematics of stars in the Galaxy provide information about its
structure and formation history. Early Galaxy formation
models, most notably those by Eggen et al. (1962),
postulated a monolithic collapse of gas and dust, with a large cloud
of gas collapsing and forming stars. Later, \citet{sz78} suggested
that the Galaxy was formed through a series of accretion events, with
smaller proto-galaxies coming together to form a larger
structure. Their scenario is more in keeping with our current
understanding of how structures formed in the early universe in a
``bottom-up'' fashion \citep{sm94,ws00}.

Ibata et al. (1994) discovered that the Galaxy is currently accreting
the Sagittarius dwarf spheroidal galaxy. Stars from this system are
not yet well-mixed with the rest of the Galactic halo either spatially
or in velocity space. Could other halo accretion events be detected if
the stars from the accreted object have become well-mixed with the
stars in the halo spatially, but not in velocity or angular momentum
space?  \citet[][hereafter H99]{h99}, found evidence in angular
momentum space for streams in the solar neighborhood using a sample of
97 metal-deficient red giants and RR Lyrae stars within 1 kpc of the
Sun and, using the sample properties, generated a model of the
streams.

The purpose of this work is to search for further evidence for
substructure in the local Galactic halo and to develop novel ways to
detect halo substructure. In order to detect the subtle signs of
kinematical substructure, accurate data are needed. We use two
different halo star samples: the original H99 sample and a combined
data set that includes local giants, red horizontal branch (RHB)
stars, and RR Lyrae stars, all selected without kinematic bias. Both
samples have full space velocities, distances, and
metallicities. Further improvements to the solar neighborhood halo
sample will be presented in the next paper in this series,
H.L. Morrison et al. (2007, in preparation, hereafter M07), which will
focus on the overall properties of the sample and their constraints on
the formation history of the local halo, rather than specifically on
substructure.

Full space velocities are difficult to obtain for many stars,
preventing a complete analysis of their motions like that of
H99. However, the H99 models and original data show structure even in
their radial velocity distributions, especially when stars in a
particular direction, e.g. near the Galactic poles, are considered. In
addition, other possible substructures that we discuss in this paper
occupy the tails of the angular momentum distribution, and so may be
detectable in radial velocities dominated by rotational velocity.
Standard statistical tests such as the Shapiro-Wilk test for
normality, which is quite sensitive to the behaviour of the tails of
the distribution, can be used to detect such deviations. We use the
H99 model of the streams as well as a model of a smooth halo to show
how streams can be detected in one component of velocity ($v_z$), and
we determine the detection limits for this method. We also extend this
method to radial velocities for stars in the direction of the Galactic
poles.

In recent years, increasingly large and precise data sets of
abundances for multiple elements have made it possible to study a new
dimension of substructure in the halo: the chemical
patterns. Abundances trace the star formation history of the objects
that became today's halo.  A good general discussion of the
possibilities of this technique is given in \citet{kcf_josh}.  In
particular, the ratios of the $\alpha$-elements such as Mg, Ca, and Ti
to Fe allow the possibility of distinguishing between a chemical
history where enrichment only comes from short-lived massive stars
(Type II supernovae), and one where the contribution of Type Ia
supernovae need to be considered as well. Type II supernovae
preferentially enrich the interstellar medium (ISM) with $\alpha$
elements, while Type Ia supernovae produce more iron but need a longer
period of chemical evolution. Thus the bulk of the local halo stars
have a higher value of [$\alpha$/Fe] than local disk stars, reflecting
different formation conditions. However, there are some metal-poor
stars that have unusually low values of [$\alpha$/Fe], such as those
in dwarf spheroidal galaxies \citep{shetrone01,shetrone03,tolstoy03},
a small number of outer halo globular clusters such as Pal 12 and Rup
106 \citep{brown97,brown99}, and some local halo field stars
\citep{carney97,king97,nissen97,fulbright2002,stephens2002,inese03,venn04}. For
stars from substructures identified via kinematic methods, we
investigate whether there are any unusual patterns in [$\alpha$/Fe].

Section \ref{datasets} describes the compilation of the data sets used
in this work in detail. Information about our model of the Galactic
halo is given in Section~\ref{model} and our results are described and
discussed in Section~\ref{results}. Section~\ref{conclusions}
summarizes our conclusions.

\section{Data Sets} \label{datasets}

The search for substructure in H99 is particularly powerful because
estimates of all three space velocities are available for their
sample. This allows the use of angular momentum (approximately
conserved in a roughly spherical potential) to isolate stars with
similar origins. In the pre-{\em Gaia} era, the need for three space
velocities limits us to a relatively small volume surrounding the Sun,
where sufficiently accurate proper motions are available.  It is
important to note that methods of identifying satellite debris using
physical quantities such as angular momentum and energy
\citep[e.g.][]{lb2,h99,h06} rely strongly on good distance estimates.
(The angular momentum estimate varies as distance squared, since two
out of three velocity coordinates are obtained by multiplying the
proper motion by the distance.) Thus the accuracy of distance
estimates is very important.

\subsection{H99 Data}

The H99 sample stars were selected from the compilations of
\citet{bs95} and \citet[][hereafter CY98]{cy98} to have distances less
than 1 kpc and [Fe/H] less than --1.6.  This sample contains nearby
stars in three evolutionary states: first ascent red giants, RHB
stars, and RR Lyrae variable stars. These stars present different
challenges for stellar population work. Good velocity estimates are
difficult to obtain for RR Lyrae stars, but are relatively
straightforward for red giants. This is because RR Lyrae stars show
significant velocity variation due to pulsation, while first ascent
giants and RHB stars have spectra that lend themselves readily to
accurate velocity estimation. Conversely, distance estimates for RHB
and RR Lyrae stars are good, but distance estimates for red giant
stars less so.  RR Lyrae and RHB stars show little variation in
absolute magnitude with metallicity. For example, \citet{mm98}
estimated a typical error of 10\% for RR Lyrae distances, and
\citet{vivas05} estimate 6\%. First ascent red giants generally have
less precise distances, because the position of the giant branch in
the color-magnitude diagram depends strongly on metallicity, and so
metallicity measurement errors propagate to a larger distance error of
order 20\% (see discussions in \citealt{mff90,mo03}). The advantage of
smaller distance errors for RHB stars is offset by the difficulty of
identifying these stars; in order to separate the subtle differences
in gravity between first ascent red giants below the horizontal branch
and horizontal branch stars we need photometry from intermediate-band
systems like Stromgren or DDO \citep{b80,nbp,att94}. Luckily, many of
the stars in the H99 sample were observed by \citet[][hereafter
ATT94]{att94} who obtained accurate Stromgren {\it uvby} colors,
enabling them to classify stars as either RHB or first-ascent
giants. Below we calculate the actual errors on distances, space
velocities and angular momenta for the ATT stars, illustrating that
these quantities are known remarkably precisely for a sample including
red giants.

\subsection{Combined Data Set} \label{combined_sample_sec}

Since H99 was published there have been several improvements on the
information available for the solar neighborhood giant and horizontal
branch stars. In particular, \citet[][B00 hereafter]{b00} presented a
catalog of 2016 stars selected without kinematic bias, and we use
preliminary results from a new sample of RR Lyraes with significantly
improved radial velocity measurements (M07).

\subsubsection{RR Lyrae Variables}

While the \citet{l94} data set provided accurate [Fe/H] and distance
measurements for the local RR Lyrae stars, the velocity measurements
were less accurate (typical errors of 30~km~s$^{-1}$).  M07 obtained
more accurate velocities for the RR Lyrae stars accessible from the
north, observing most stars more than once and using high-quality
light curves obtained by T.D.  Kinman (1991, private communication) to
correct for the large radial velocity amplitude of each
star. Velocities were corrected for pulsation using the synthetic
velocity curves of \citet{liu}. Typical velocity errors are
15~km~s$^{-1}$. \citet{mm98} published improved values of proper
motion for RR Lyrae stars by averaging {\em Hipparcos} proper motions
with accurate ground-based determinations from the USNO ACT reference
catalog \citep{ucw98}. We have selected a sample of 96 stars with
[Fe/H] less than --1.0 from these data sets. Metallicities for these
stars are from \citet{l94,l96}. The distances of stars in the M07
sample (derived using the \citet{l94} period-luminosity relation)
range from about 0.5 kpc (RR Lyrae itself) to 2.5 kpc. Our sample
presently only includes RR Lyrae stars of type ab.


\subsubsection{Red giants and RHB Stars} \label{combined_rg}

We have selected a subset of the red giant stars in the B00 catalog
for our expanded sample. In order to preserve the distance accuracy so
important to the calculation of angular momentum, we critically
compared the distance estimates given in the B00 catalog with those
given by earlier work such as ATT94. This uncovered some surprisingly
large, systematic differences in distance estimates.

\begin{figure}
\centering
\includegraphics[scale=0.65,angle=-90]{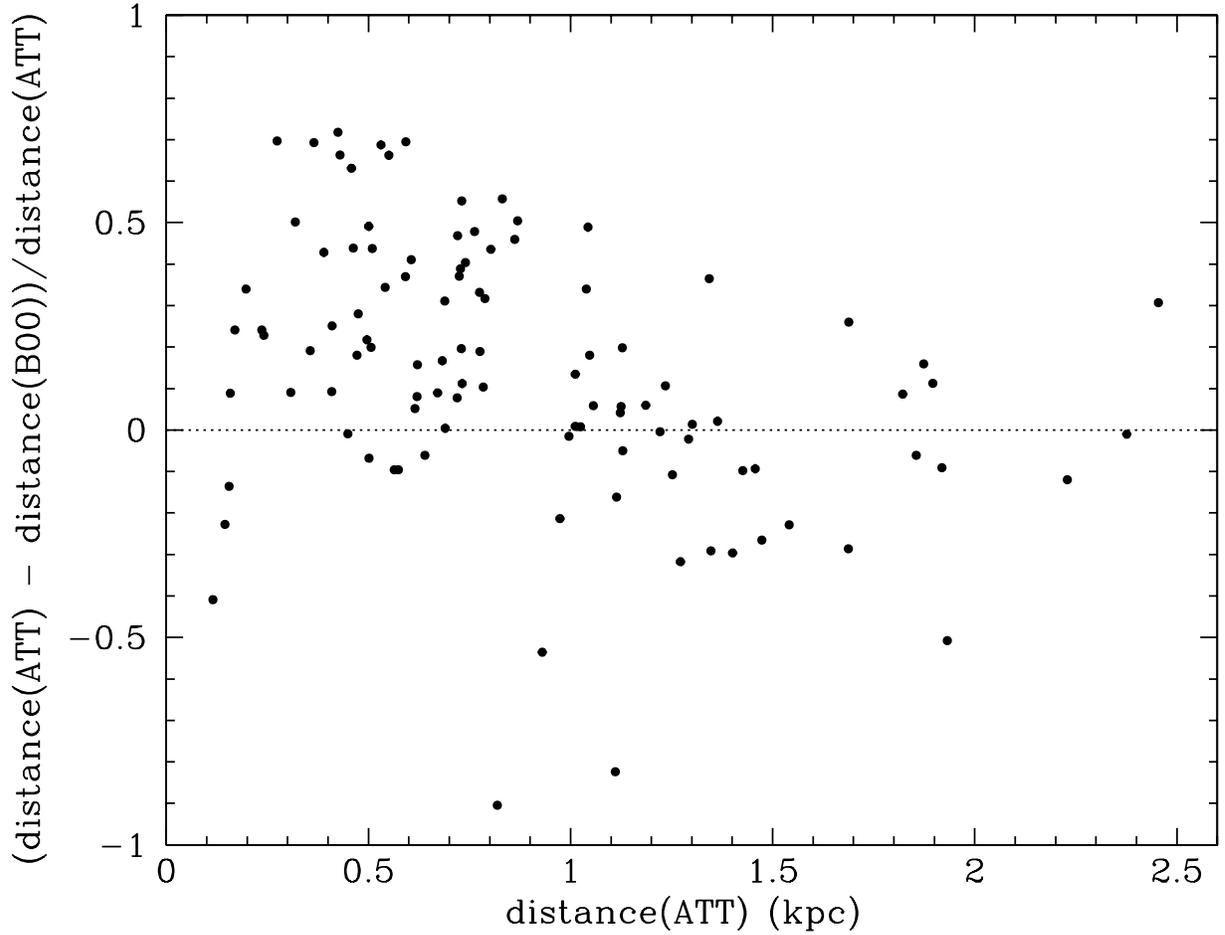}
\caption{Difference between distance estimates for red giants from ATT94
and B00 as a function of distance. The differences are expressed as a
fraction of the ATT94 distance. It can be seen that for stars closer
than 1 kpc, the B00 distances are on average 40\% smaller than the ATT94
distances.}
\label{boo_cy_dist}
\end{figure}

Figure~\ref{boo_cy_dist} shows that the B00 and ATT94 distances agree
reasonably well for distances greater than 1 kpc. For distances less
than 1 kpc, the ATT94 distances are on average a factor of 2 larger
than the B00 distances, a remarkably large amount for local,
well-studied stars.

\begin{figure}
\centering
\includegraphics[scale=0.70,angle=-90]{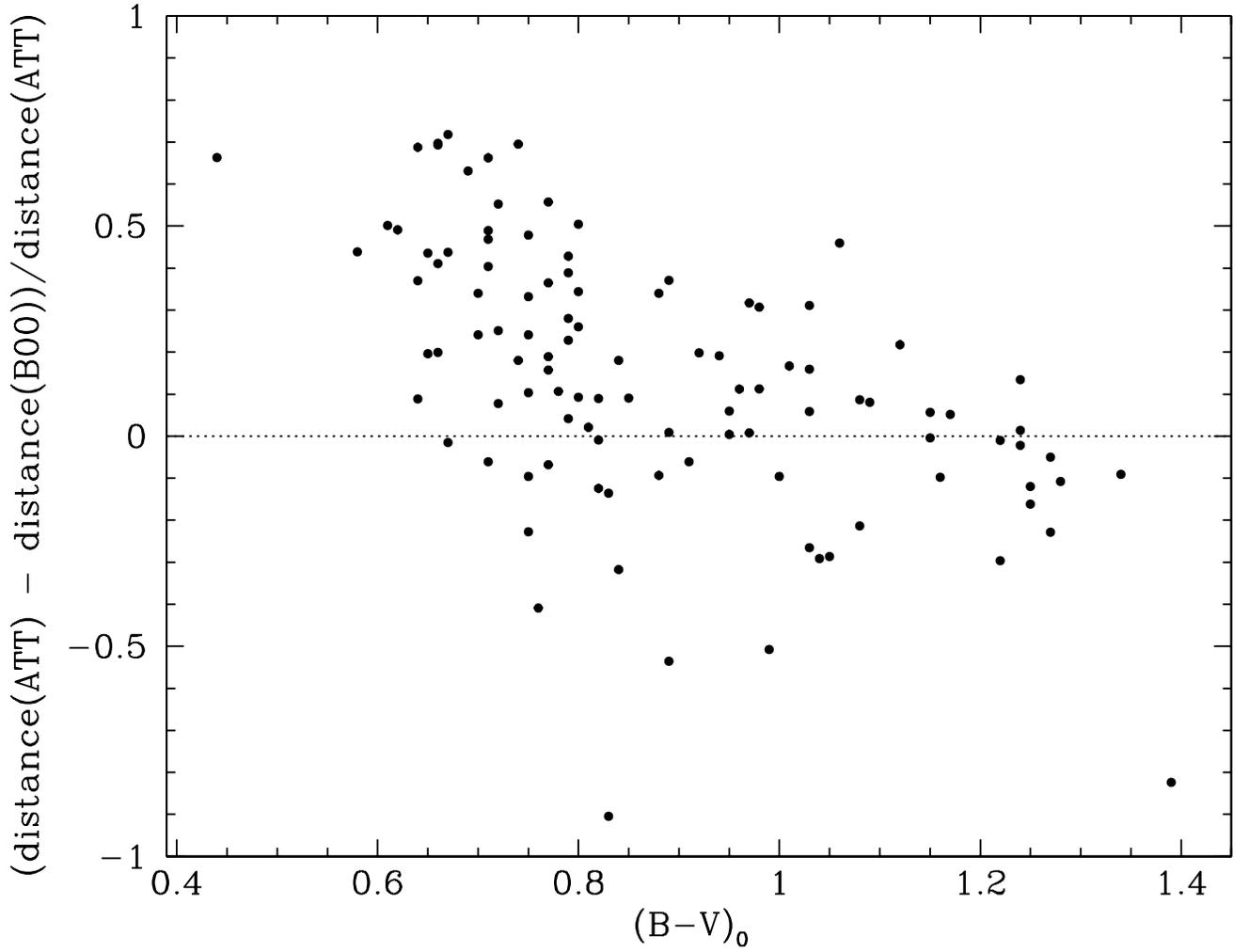}
\caption{Difference between red giant distance estimates from ATT94 and
B00, shown as a function of (B--V)$_0$ color. It can now be seen that
the problematic stars are the bluest ones, with (B--V)$_0 < 0.9$.}
\label{boo_cy_bv}
\end{figure}

Both papers use \bv\ colors and the globular cluster giant branch
parameterization of \citet{nbp} to estimate absolute magnitudes for
the giant stars. B00 assumed that all giants are on the first-ascent
giant branch, while ATT94 used the Stromgren $c1$ index to separate
horizontal-branch and asymptotic giant branch (AGB) stars from
first-ascent giants.  AGB star contamination is expected to be small
because of the short AGB lifetimes, but horizontal-branch stars are
more common. Blueward of $(B-V)_0\simeq0.9$, we expect significant
numbers of RHB stars to appear in the sample. Assuming that these
stars were on the first ascent giant branch, as B00 did, will lead to
significant underestimates of their distance. This error will become
larger for the bluest stars. Figure \ref{boo_cy_bv} shows that this is
indeed the case, with B00 distances being on average a factor of 2
smaller than the ATT94 distances for $(B-V)_0=0.6$.

Thus we have restricted our sample to giants from B00 with $(B-V)_0$
greater than 0.9, red giants and RHB stars from CY98 with distances
determined by ATT94, and RR Lyrae stars from the M07 sample.  We have
also restricted the sample to stars with [Fe/H] less than --1.0 and
distances less than 2.5 kpc, and eliminated stars with thick disk
kinematics (see Section~\ref{streams_in_combined} for details). The
$(B-V)_0$ criteria led us to accept 24 red giants and to reject 85 red
giants from B00 that had otherwise acceptable distances and
metallicities. Fourteen of the accepted stars were later eliminated
from the final sample because of their thick disk kinematics (see
Section~\ref{streams_in_combined}).  For stars in our sample within
0.5 kpc of the Sun, our distances agree with {\em Hipparcos} distances
\citep{hipcat} within the errors. The only exception is HD 135449,
whose distance in Table~\ref{combined_sample_table} is twice the
distance given by Hipparcos. In future versions of this sample
(M07), we plan on using the
Hipparcos distances for nearby stars. Metallicities in B00 come from a
large variety of sources, but the ATT94 data (which dominate our
combined sample of red giants and RHB stars) are significantly more
homogeneous, being based either on {\it uvby} photometry (89 stars, an
estimated [Fe/H] error 0.16 dex) or accurate spectroscopic abundances
from the literature (68 stars).


\subsubsection{Error Estimates}

Because the data in our combined sample come from a variety of sources
and include stars in different evolutionary stages, it is preferable
to calculate the errors on derived quantities individually, since they
can vary significantly (see Table 1 and Fig.
\ref{meta_fehm10_d25_ang_mom_groups}).  We propagated the known errors
on distance, velocity, and proper motion through the calculation of
$v_R, v_\phi, v_z$, $J_z$, and $J_{\perp}$ using a Monte Carlo
calculation, drawing new values of the input quantities from Gaussian
distributions, recalculating the derived values and measuring the
standard deviation of their distributions.

For the RR Lyrae and RHB stars, distance errors are relatively small,
and have been assumed to be 7\%. For the first ascent red giants in
the sample, it is more difficult to quantify errors because the
metallicity is the main driver of distance errors and many authors do
not quote metallicity errors\footnote{Note that CY98 calculated
distance errors simply by assuming that the error on each metallicity
value in their sample was 0.16 dex, while ATT94's actual quoted errors
vary from 0.01 to 0.31 dex. We have set errors quoted at less than 0.1
dex to be 0.1 dex in our calculations.}. We have calculated distance
errors individually for stars in the ATT94 sample (where accurate
metallicity errors are given, and in fact a significant number of
stars have multiple high-dispersion metallicity estimates) and for a
small number of red giants, most of which are part of the outlier
groups discussed in Section~\ref{streams_in_combined}. To calculate
distance errors, we used \bv\ color and the globular cluster giant
branch loci of \citet{nbp} with a Monte Carlo calculation of the
effect of metallicity and color errors on the estimate of absolute
magnitude.

For the five red giant stars not in the ATT94 sample, we made the
following assumptions. Four (HD 18710, HD 174578, HD 214925 and CD
--68 1881) have [Fe/H] values from \citet{nbp}. \citet{tat94} have
shown that the DDO [Fe/H] calibration used for these stars has
systematic errors in some metallicity ranges. We have assigned [Fe/H]
errors of 0.25, 0.50, 0.20 and 0.50 dex respectively to account for
this. BD +30 2282 has metallicity from \citet{hy82}; we have assigned
an [Fe/H] error of 0.35 dex for this star. There was also one giant in
ATT94 (HD 128279) that was too blue to use the \citet{nbp} distance
calibration for the Monte Carlo calculation.  Luckily, it has a {\em
Hipparcos} parallax of reasonable accuracy so we have simply used the
parallax and its error rather than our other distance error estimation
procedure.

\subsubsection{Final Sample}

We have error estimates for distances and derived quantities
(velocities and angular momentum) for the great majority of the
sample: 210 out of 231 stars.  The combined sample has a median
distance of 1.1 kpc, median distance error 7\%, and median metallicity
[Fe/H] = --1.7. Median errors on $v_R, v_\phi, v_z$ velocities are 15,
20 and 11 km~s$^{-1}$ respectively, making this sample well-suited for
careful investigations of substructure.

Table \ref{combined_sample_table} lists values of distances,
metallicities, radial velocities, and galactocentric velocities for
the red giants and RHB stars in the combined sample as well as the
associated errors for these quantities. The RR Lyrae data will be
published in M07.  In Table~\ref{combined_sample_table}, the radial
velocities are heliocentric and the $v_R$, $v_\phi$, and $v_z$
velocities are relative to the center of the Galaxy on a left-handed
coordinate system. This has $R$ pointing away from the center, $\phi$
increasing in the direction of Galactic rotation, and $z$ pointing
towards the north Galactic pole (NGP). Note that our angular momenta
are calculated on this left-handed system in order to directly compare
our results with the results of H99. (See the Appendix for more
details.)  The velocities were corrected for the motion of the Sun and
LSR using $v_{lsr} = 220.0 \ \kms$, $U_{\odot} = -9.0 \ \kms$,
$V_{\odot} = 12.0 \ \kms$, and $W_{\odot} = 7.0 \ \kms$ on the
standard left-handed coordinate system \citep{blaauw,mb81}. The Sun
was assumed to be 8.0 kpc from the Galactic center. See the Appendix
for the transformation between the local, Sun-centered coordinate
system and the galactocentric coordinate system.

\begin{figure}
\centering
\includegraphics[scale=0.7,angle=-90]{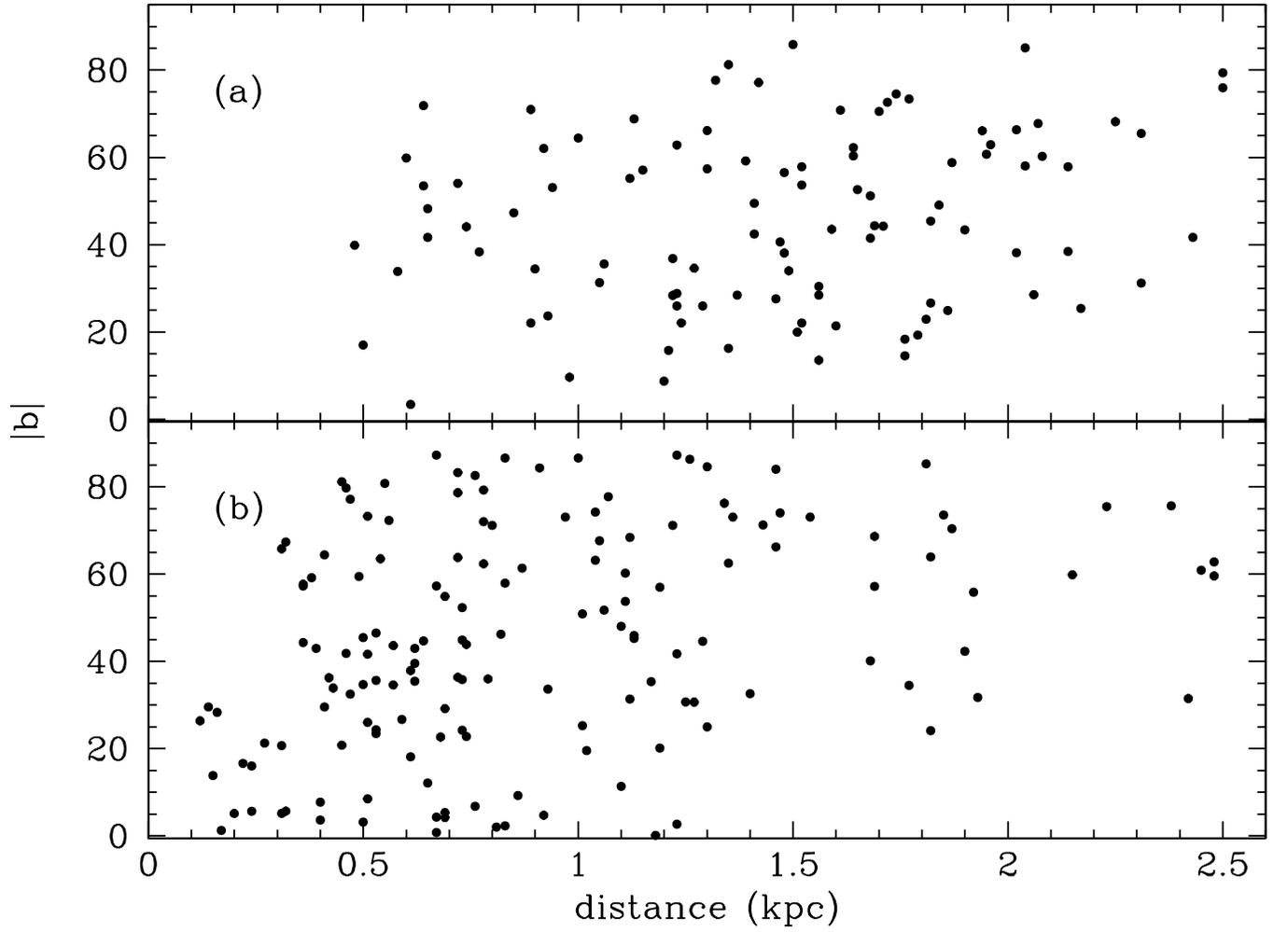}
\caption{Distribution of distance and absolute value of Galactic
latitude ($b$) for (a) the RR Lyrae variables in our sample and (b) the
red giant and RHB stars in our sample.}
\label{lb}
\end{figure}

The sample was selected without kinematic bias, but is spatially
incomplete: it has fewer stars with low $|b|$ because of the
difficulty of identifying these rare halo stars in heavily reddened
regions. We show the distribution of distance and $|b|$ in
Figure~\ref{lb}. It can be seen that the RR Lyrae sample is more
complete in $|b|$ than the giant sample, although neither are fully
complete in $|b|$. The giant sample has a lower mean distance than the
RR Lyrae sample because of the intrinsic rarity of the RR Lyrae
stars. Since all three velocity components can be measured for each
star, the spatial incompleteness does not cause a bias in the velocity
distribution directly. However, the bias toward high Galactic latitude
does select against stars with a flattened space distribution, such as
the metal-poor stars in the thick disk \citep{nbp,mff90}. Therefore,
we would expect any such stars to be under-represented in our sample.

Since there will be a significant number of binaries in any sample of
stars (for example, 17\% in the sample of metal-poor giants of
\citealt{carney03}), we need to consider the effect of binarity on our
estimates of distance and velocity.  Fortunately, because our sample
is composed of giants and horizontal branch stars, the effect of any
undetected binaries on the distance estimates will be
small. \citet{carney03} studied 91 metal-poor red giants and RHB
stars, (68 in common with our sample) and identified 8 spectroscopic
binaries. The median velocity amplitude of these binaries is
8~km~s$^{-1}$, which is small compared to our typical error on a U,V
or W velocity (10-15~km~s$^{-1}$). While we expect a similar number of
undetected spectroscopic binaries in our sample, the effect on our
conclusions will be negligible. Similarly, the ``velocity jitter''
that is seen in some stars close to the giant branch tip is of an even
smaller magnitude (of order 5~km~s$^{-1}$) and will not affect our
conclusions either.

\section{Model Information} \label{model}	

\begin{figure}
\centering
\includegraphics[scale=0.7]{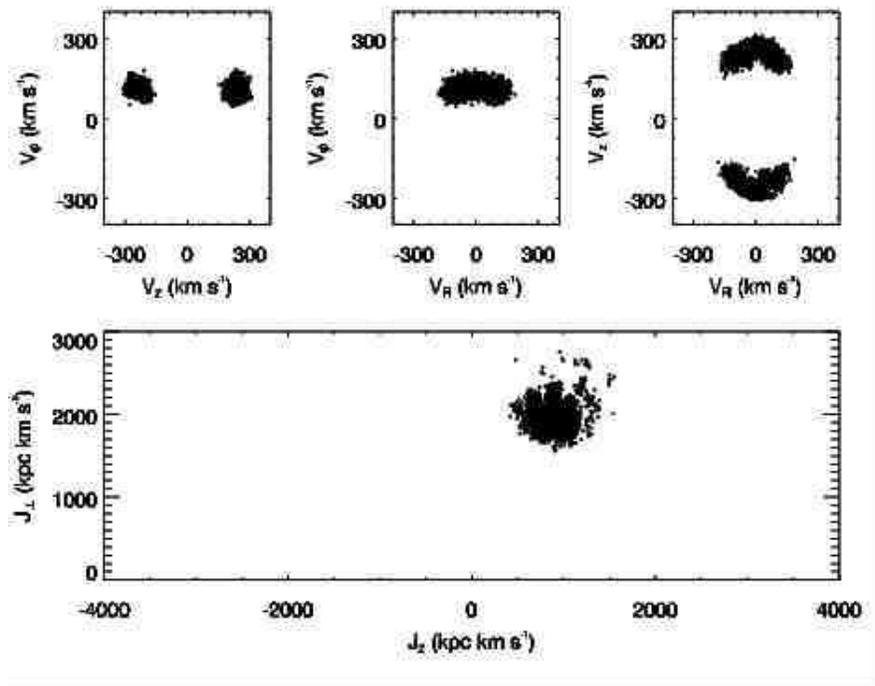}
\caption{Structure in velocity and angular momentum space for the H99
  star streams model for stars within 2.5~kpc of the Sun. The bottom
  panel is a plot of the angular momentum in the plane of the Galaxy's
  disk as a function of angular momentum out of the disk. The top
  panels show the cylindrical velocity coordinates of the data plotted
  against each other.}
\label{ang_mom_am}
\end{figure}

\citet{hw99} described models used to study the disruption of satellite
galaxies by the Milky Way. H99 used these to model the progenitor of
the star streams they detected in the solar neighborhood. The remnants
of a progenitor with an initial velocity dispersion of 18~km s$^{-1}$
and core radius of 0.5-0.65~kpc fit the star streams discovered in
H99. Its orbit has an apocenter at 16~kpc, pericenter at 7~kpc, a
maximum height above the plane of 13~kpc, and a radial period of
0.4~Gyr. The present paper uses an improved version of the model,
which better accounts for the self-gravity of the satellite and has
positions and space velocities for $10^6$ particles in the disrupted
progenitor from 0~Gyr-13~Gyr.  After 3-5 Gyr, these particles
fill much of the space between 7 and 16 kpc from the Galactic center
(the amount of debris from the progenitor peaks at the solar circle)
and are very well-mixed spatially within $\sim$5 kpc from the Sun.
The disrupted progenitor still has some structure in velocity and
angular momentum space, however, as can be seen in
Figure~\ref{ang_mom_am}.

\begin{figure}
\centering
\includegraphics[scale=0.7]{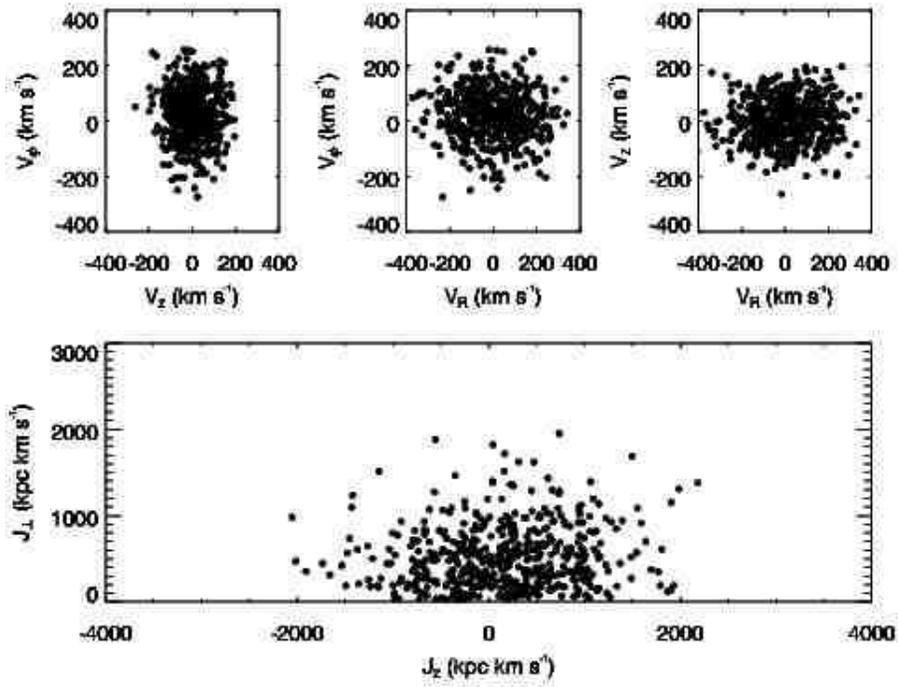}
\caption{Structure in velocity and angular momentum space for the
smooth model for stars within 2.5~kpc of the Sun.}
\label{ang_mom_smooth}
\end{figure}

In addition, we have created a model of a smooth halo. This model
allows us to test the null hypothesis that the halo has no
substructure. The density distribution of the smooth halo (number of
stars per unit volume) is proportional to $r^{-3.5}$ \citep{z85,vivas06}. The
velocity distribution of the halo was modeled using a velocity
ellipsoid \citep{swar}.  The models were created by randomly selecting
points from these distributions. Figure~\ref{ang_mom_smooth} shows one
realization of this model in velocity and angular momentum space. The
parameters $\avevR$, $\avevphi$, $\avevz$, $\sigma_R$, $\sigma_\phi$,
and $\sigma_z$ were determined by finding the average and the standard
deviation of the $v_R$, $v_\phi$, and $v_z$ velocity distributions
from the combined data set excluding known stream stars.

The standard velocity ellipsoid numbers for the halo
\citep[e.g.][]{cb00} were not used because the H99 star streams, in
particular, bias the velocity distributions. For example,
Figure~\ref{ang_mom_h99} shows that the stream stars have large
absolute $z$ velocities. Including these stars in the velocity
ellipsoid calculations changes $\sigma_z$ from 84 km s$^{-1}$ to 101
km s$^{-1}$.   Table~\ref{smo_params} gives the derived velocity
ellipsoid parameters for our combined sample (stars with distances
less than 2.5 kpc and $\feh \leq -1.0$, excluding likely thick disk
stars), both with and without stream stars.
In the case of $\avevR$ and $\avevz$, the parameters used are
consistent with zero within the errors. However, $\avevphi$ is not
consistent with zero. \citet{cb00} find a similar $\avevphi$ for their
sample within 1~kpc of the Sun and note that $\avevphi$ decreases when
larger distance ranges are included. Note that excluding the star
streams from the sample used to determine the velocity ellipsoid also
increases the anisotropy of the halo.


\section{Identifying Structure in the Tails of the Velocity Distribution} \label{results}

\subsection{Method} \label{method}

\begin{figure}
\centering
\includegraphics[scale=0.7]{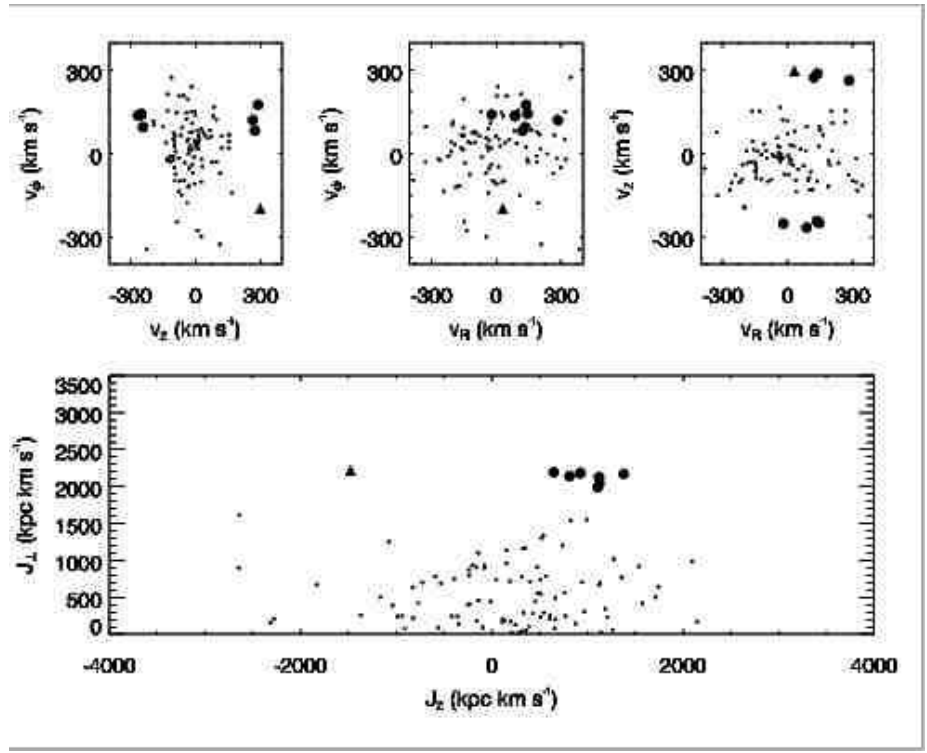}
\caption{Plots of the distribution of the H99 data in velocity and
angular momentum space for stars with $\feh \leq -1.6$ and distances
less than 1.0 kpc from the Sun.  The
stream stars are represented by the large circles. The additional
$v_z$ outlier (HD 124358) is shown by the upward pointing triangle. }
\label{ang_mom_h99}
\end{figure}

Using distances and full space velocities of metal-poor red giants and
RR Lyrae stars, H99 identified the remnant of a merger between a small
satellite galaxy and the Milky Way. Since obtaining proper motions of
stars is difficult, another way to identify subtle merger remnants is
desirable. Here we explore signatures visible in the tails of the
velocity distribution. Figure~\ref{ang_mom_h99} plots the H99 data in
velocity and angular momentum space. Its bottom panel plots $J_z$, the
component of the angular momentum (per unit mass) out the plane of the
Galaxy's disk, versus $J_{\perp} = \sqrt{J_x^2 + J_y^2}$, the angular
momentum (per unit mass) in the plane of the disk. The H99 stream
stars ({\em large circles}) clearly occupy a different region of phase
space than the rest of the halo stars, attesting to the debris' common
origin as a Galactic satellite. Figure~\ref{ang_mom_h99} ({\em top
panels}) shows cylindrical velocity coordinates.  The stream stars
({\em large circles}) are located in two clumps, one with $v_z \sim
-200$~km~s$^{-1}$ and the other with $v_z \sim 200$~km~s$^{-1}$. The
bimodality in $v_z$ seen for the nearby stream stars comes about
because their orbits reach to a large distance (13 kpc) above and
below the Galactic plane. Therefore, any nearby substructure stars
will be passing through with high positive or negative $v_z$
values. We can anticipate that this bimodality will be less marked as
more distant stars are included in the sample, because it will include
stream stars with smaller values of $|v_z|$.

\begin{figure}
\centering
\includegraphics[scale=0.7]{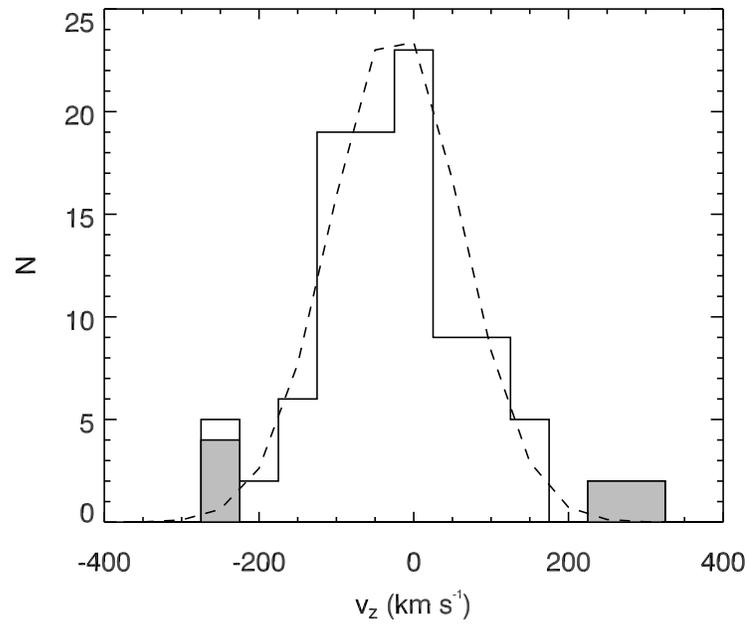}
\caption{Histogram of the H99 data $v_z$ velocity distribution with
  the stream stars shaded. A normal distribution is plotted as a
  dashed line. Note how the H99 stream stars widen the wings of the
  velocity distribution making it deviate from normality.}
\label{vz_hist_h99}
\end{figure}

Note that the $v_z$ components of the H99 streams extend well into the
tails of the velocity distribution, as can be seen in
Figure~\ref{vz_hist_h99}.  A smooth halo velocity distribution is
well-approximated as a multi-dimensional normal (Gaussian)
distribution \citep[e.g.][]{bm98}, so testing the $v_z$ distribution
for normality may reveal the presence of substructure like that in
H99. In addition, the line of sight component of velocity for stars
near the Galactic poles is dominated by the $v_z$ component of the
star's velocity.  Therefore, if one uses a sample of stars near the
Galactic poles, one should be able to test for structure in the $v_z$
distribution using only radial velocities. While with this technique
we are only able to identify the presence of substructure in the data,
not necessarily the substructure itself, it will be useful in
identifying data sets where other observers might want to aggressively
pursue proper motions and distances.\footnote{One might imagine
employing a genetic-algorithm technique to find stream stars where
random groups of stars are selected, the Shapiro-Wilk product
calculated, and the groups that have the highest p-values are bred
together to create the next generation of groups to test. This is, of
course, a computationally intensive method.}

There are many tests for deviations from normality; one of the most
powerful is the Shapiro-Wilk test \citep{sw65,da86}.  This test is
based on the concept of probability plots, which plot the cumulative
distribution function\footnote{The cumulative distribution function
gives the fraction of data points that are less than or equal to a
value as a function of value: $CDF(x) = P(X \le x)$, where $x$ is the
value and $X$ is the data.} using a transformation of the vertical
axis that makes normally distributed data fall along a straight line.
Examples of probability plots are shown in Chapter 1 of
\citet{daste86}.  The slope of this line gives an estimate of $\sigma$
for a normal distribution.  The Shapiro-Wilk test compares this slope
with the sample standard deviation \citep{st96}. This technique is
particularly sensitive to deviations from normality in the tails of
the distribution. Code for this test \citep{ro95} can be obtained at
StatLib.\footnote{http://lib.stat.cmu.edu/apstat/R94}


To test this method, we applied the Shapiro-Wilk test to the H99 $v_z$
data. Here the null hypothesis is that the data have a normal $v_z$
distribution, and the probability value (p-value for short) indicates
how often data with a normal distribution would produce this
data set. Thus the smaller the p-value, the less likely it is that the
data come from a normal distribution.  (Note, however, that a high
p-value does not mean that the sample distribution is normal, just
that it is consistent with a normal distribution.) The results of this
test are given in Table~\ref{sw_h99}. We list the tested data set and
the associated p-value. The entire sample fails the Shapiro-Wilk test
at the 1\% level. When the H99 stream stars and one other star (HD
124358) at ($J_z \sim -1500$, $J_\perp \sim 2300$) are excluded,
however, the sample passes the Shapiro-Wilk test. The additional
excluded star, indicated by a large triangle, has a very retrograde
orbit and is well away from the rest of the data in angular momentum
space; it may belong to another stream (see
Section~\ref{streams_in_combined}). These results demonstrate the
effectiveness of this method for detecting streams like those in H99,
where the stars occupy the tails of one component of the velocity
distribution.

\subsection{Modeling Stream Detections with Distance} \label{modeling}

Having established that we are able to detect substructure using the
$v_z$ velocity distribution, now we use the H99 model of the debris
and our model of the smooth halo to see (1) how changing the sample
distance limit affects our ability to detect streams and (2) how close
stars need to be to the Galactic pole to detect substructure using
radial velocity information only.

\begin{figure}
\centering
\includegraphics[scale=0.7]{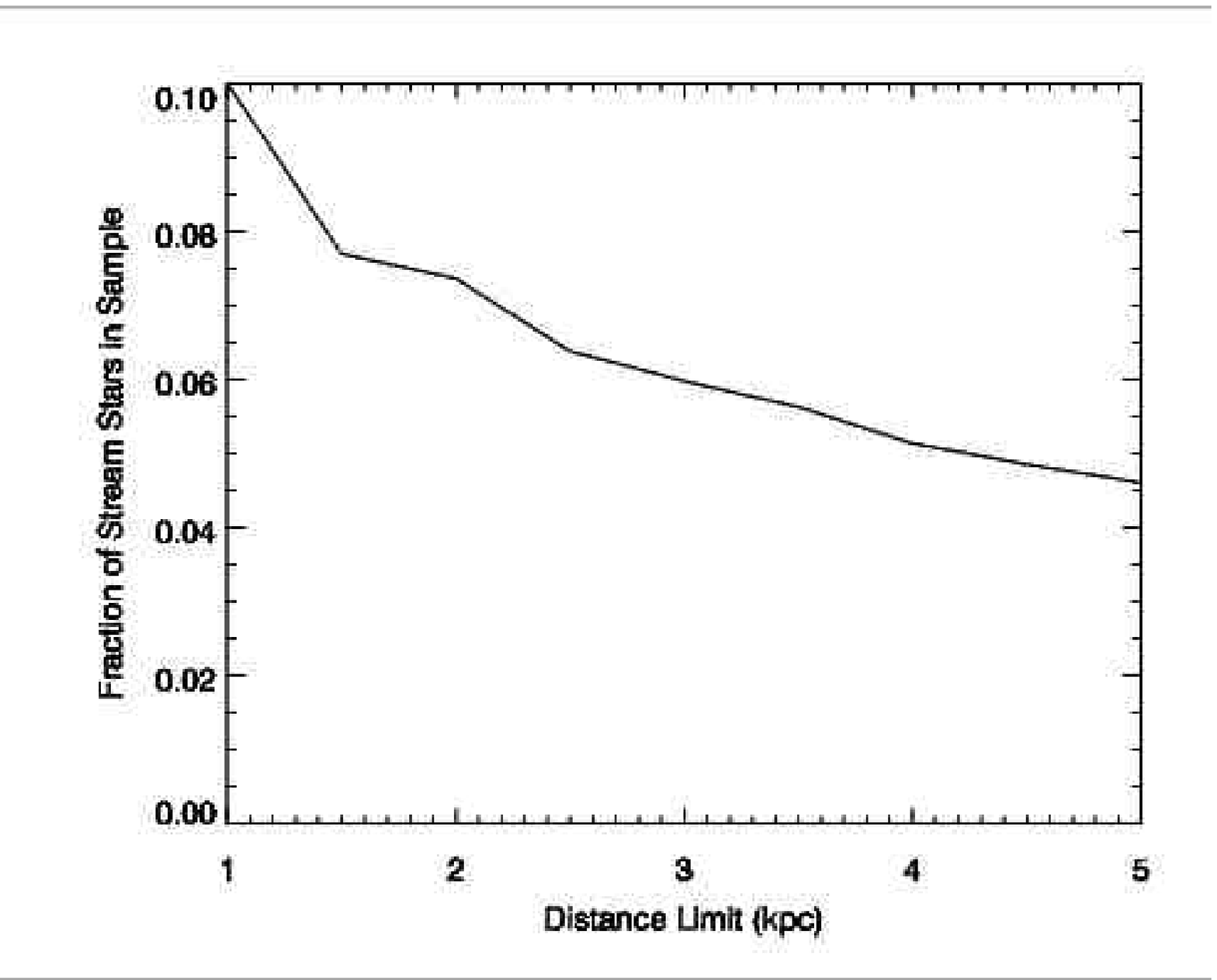}
\caption{Fraction of stream stars in halo as a function of distance
limit. Note that the fraction of stream stars decreases as the
distance limit increases when the fraction of stream stars is
normalized to 10\% of the halo stars within 1~kpc of the Sun.}
\label{frac_stream_plot}
\end{figure}

To characterize stream detections, we produced random samples with
sizes ranging from 20 to 4000 stars, and distance limits of 1, 2, 3, 4
and 5 kpc.  In each case, we produced 10,000 different realizations of
the sample.  While the stream stars are spatially well-mixed up to 5
kpc from the Sun, the relative number of stream stars to smooth halo
stars drops slowly as we move away from the Sun; see
Figure~\ref{frac_stream_plot}. The number of stream stars in each
distance-limited sample was scaled using the models so that stream
stars made up 10\% of the sample for a distance limit of 1.0~kpc. The
maximum distance limit in our simulations was 5~kpc, since the
velocity ellipsoid of the halo may change beyond this distance
\citep[e.g.][]{vedel}. For each sample, we tested the $v_z$
distribution for normality using the Shapiro-Wilk test. We calculated
the p-value (significance level) for each of the 10,000 realizations
of the sample, and then calculated the average p-value for each
sample.

\begin{figure}
\centering
\includegraphics[scale=0.7]{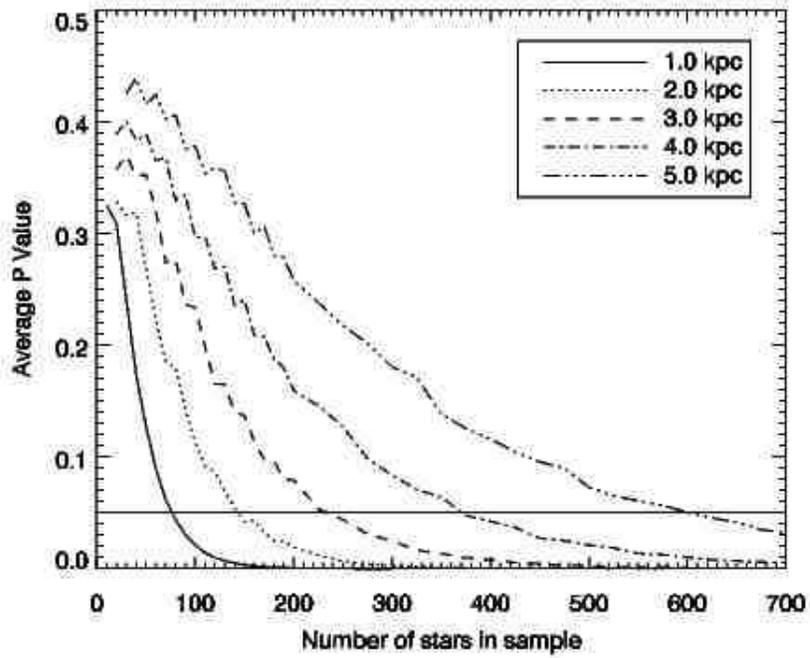}
\caption{Average p-value for 10,000 random samples of $v_z$ velocities
with the relative number of stream to smooth halo stars fixed at 10\%
for a distance limit of 1.0 kpc. The standard deviation of the trial
p-values is on the order of the width of the lines. The distance
limits used for the samples are 1.0, 2.0, 3.0, 4.0, and 5.0 kpc. As
the distance limit increases, the size of the sample for which
detections are possible also increases.}
\label{n_pw_dist_plot}
\end{figure}

Figure~\ref{n_pw_dist_plot} shows the average p-value as a function of
sample size for various distance limits in our $v_z$ samples. P-values
below 5\% fail the Shapiro-Wilk test for normality. The number of
stars needed to detect the stream with a 1.0~kpc distance limit is
less than the sample size of H99 sample with this distance limit (101
stars), so it is unsurprising that we were able to detect the H99
streams using this method (see Table
\ref{meta_fehm10_d25_vz_results}). As the distance limit increases,
more stars are needed in the sample to detect the presence of the H99
streams. For example, with a distance limit of 5~kpc, approximately
600 stars are needed to detect the H99 streams, but only 75 stars are
needed with a distance limit of 1.0 kpc. There are two causes
here. First, the fraction of stream stars in a sample decreases slowly
as the sample distance limit increases (see
Fig.~\ref{frac_stream_plot}). Second, the velocity distribution of
the stream changes away from the solar neighborhood, and this results
in fewer stream stars with extreme values of $v_z$. (Since the
high-energy debris from the progenitor is bound to the Galaxy, stars
nearest the plane will have the largest $v_z$ velocities. As the
distance from the plane increases, the $v_z$ velocities of the stars
will decrease.)  The result of both effects means that increasing the
distance limit on a sample does not necessarily increase the detection
probability, if it does not have enough stars for the stream to be
detected in the sample with the larger distance limits.

\begin{figure}
\centering
\includegraphics[scale=0.7]{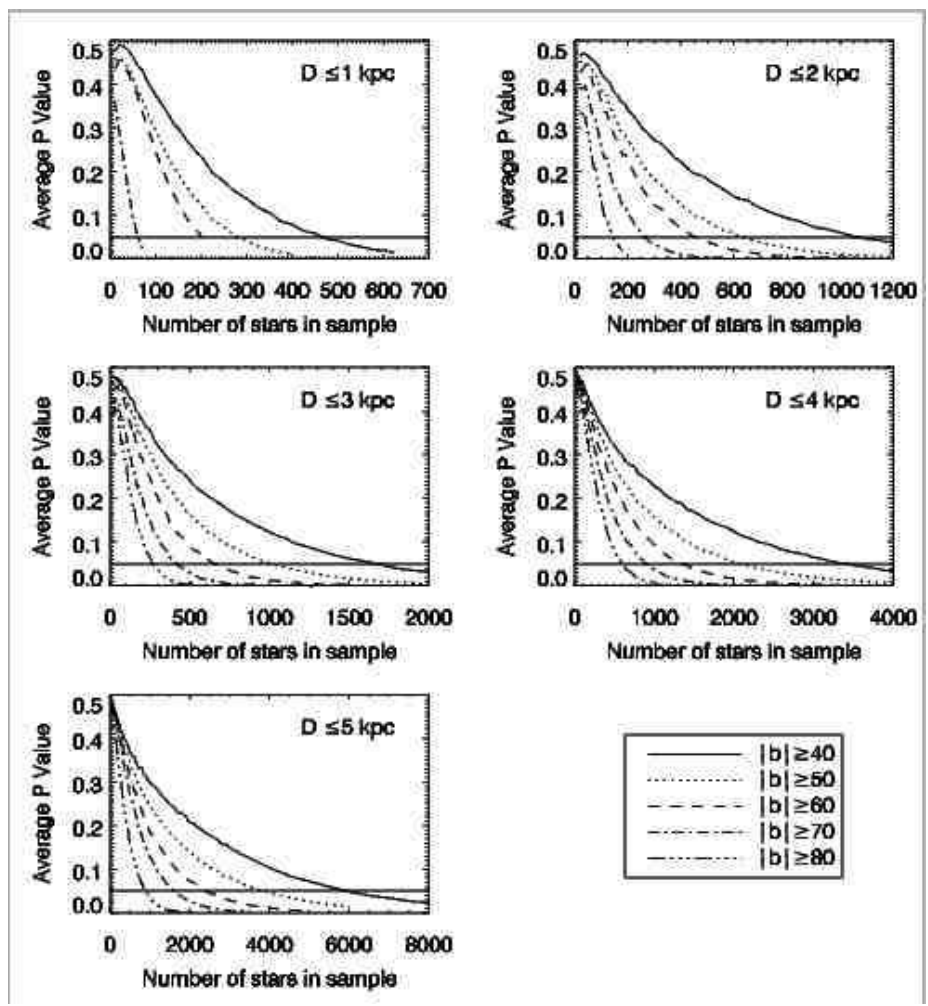}
\caption{Average p-value for 10,000 random samples of radial
velocities (corrected to the Galactocentric frame of reference) with
10\% stream stars for a distance limit of 1.0~kpc.  The standard
deviation of the trial p-values is on the order of the width of the
lines. Galactic latitude limits ($|b| \ge 40$, $|b| \ge 50$, $|b| \ge
60$, $|b| \ge 70$, and $|b| \ge 80$) were imposed on samples with five
different distance limits (1.0, 2.0, 3.0, 4.0, and 5.0 kpc). N.B. The
x-axes on these graphs all have different scales, since the range of
samples tested is very different in each case.}
\label{n_pave_dist_b}
\end{figure}

To use radial velocities 
to identify kinematic structure in $v_z$, we need to identify how
close the sample stars need to be to the Galactic poles to have enough
$v_z$ velocity information to detect structure. To examine this, we
generated five different distance limited samples ($D \le 1.0$, $D \le
2.0$, $D \le 3.0$, $D \le 4.0$, and $D \le 5.0$ kpc) and applied four
different Galactic latitude limits to each sample ($|b| \ge 40$, $|b|
\ge 50$, $|b| \ge 60$, $|b| \ge 70$, and $|b| \ge 80$).  The stream
stars were assumed to be well mixed spatially so we used the same
normalization for the stream stars as in the previous set of
simulations. We calculated the radial velocity of each star in each
sample and then used the same algorithm as above substituting the
radial velocities for the $v_z$ velocities to determine an average
p-value for the radial velocity distribution of a sample. The results
of these simulations are given in Figure~\ref{n_pave_dist_b}.

Comparing Figures~\ref{n_pw_dist_plot} and~\ref{n_pave_dist_b}, one
sees that for lower Galactic latitude limits, much larger sample sizes
are needed to detect streams. As the Galactic latitude limits get
further from the Galactic pole, the width of the smooth halo radial
velocity distribution increases, drowning out the presence of stream
stars in the wings of the distribution. As we sample closer to the
Galactic poles, the sample size needed to detect the H99 streams using
radial velocities approaches that needed to detect the streams in $v_z$
velocity space.  In short, using the radial velocity test to
overcome a lack of proper motion data requires a significant increase
in the sample size.

\subsection{Streams in the Combined Data Set} \label{streams_in_combined}

We used the combined sample described in
Section~\ref{combined_sample_sec} to investigate further the
applicability of the Shapiro-Wilk test to real data.  We concentrate
here on looking for outliers in the $J_z$ versus $J_\perp$
distribution. In future work (M07),
we will investigate structure within the $J_z$ versus $J_\perp$
distribution, i.e., within the angular momentum distribution of the
disk and the halo themselves, such as in \citet{nhf04}.

\begin{figure}
\centering
\includegraphics[scale=1.0]{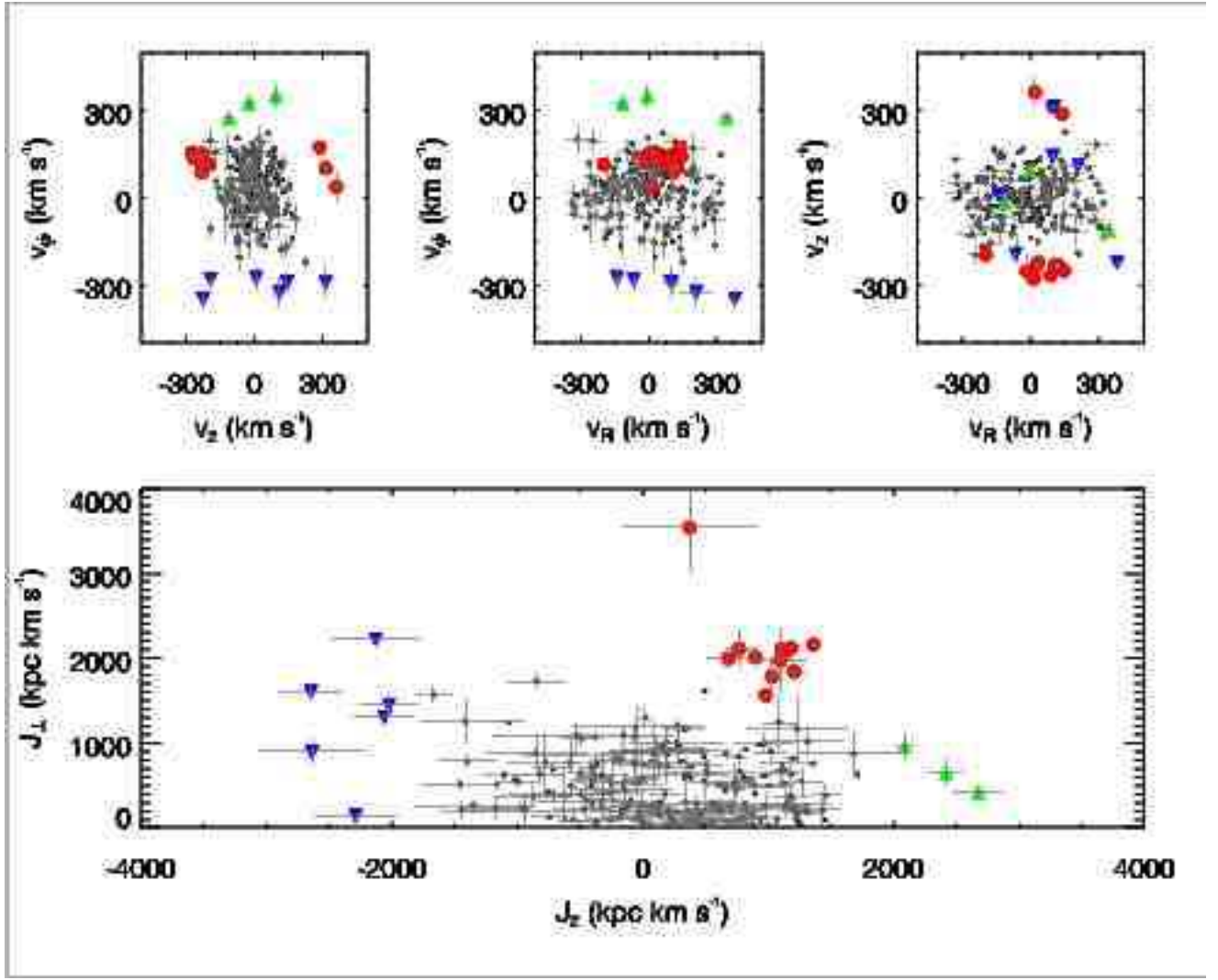}
\caption{Angular momentum and cylindrical velocity plots for the
combined data sample with the selection criteria: $ D \leq 2.5 \
\mathrm{kpc}$ and $\feh \leq -1.0$. The retrograde outliers are
indicated by downward0facing triangles, the prograde outliers by
upward-facing triangles, and the H99 streams by large circles. Note
that the empty region of the angular momentum diagram at $J_z =
1500-2500 \ \rm{kpc} \ \kms, J_\perp < 600 \ \rm{kpc} \ \kms$ was
occupied by likely thick-disk stars.  These stars were excluded from
our final sample based on their position in the angular momentum
diagram.}
\label{meta_fehm10_d25_ang_mom_groups}
\end{figure}

While H99 used two different distance limits in their analysis (1 and
2.5 kpc) in order to limit the effect of larger errors in tangential
velocity, our direct calculation of errors on the angular momenta make
this unnecessary. H99 also limited their sample to stars with $\feh
\leq -1.6$ to exclude thick disk stars. We chose simply to exclude
such stars using their position on the angular momentum plot. To be
considered a thick disk star, the star needed to have a $J_z$ between
1500 and 2500~kpc~km~s$^{-1}$ and a $J_\perp$ less than 600
kpc~km~s$^{-1}$. These limits were chosen to match the angular
momentum distribution of stars in the \citet{nordstrom04} sample with
$\feh$ greater than --1.0 and heliocentric distances less than 2.5
kpc. In addition, the star's $v_\phi$ needed to be near 220 $\kms$
with all other velocity components near zero. Note that a star
traveling at the speed ($220 \ \kms$) and position ($8.0 \ \rm{kpc}$)
of the LSR has a $J_z$ of 1760~kpc~km~s$^{-1}$. The $J_\perp$ limit
reflects the velocity dispersion of stars in the disk.  These criteria
excluded 24 stars, of which seven had $\feh < -1.6$. Note the symmetry
about $J_z \sim 0$ for the resulting halo distribution
(Fig.~\ref{meta_fehm10_d25_ang_mom_groups}). The clumpiness within the
distribution of halo stars will be investigated in M07.

We have identified the H99 stream stars and two other groups of
outliers in the combined sample. The properties of the H99 stream
stars and the outliers are given in Tables~\ref{meta_strm_table},
\ref{meta_out1_table}, and
\ref{meta_out2_table}. Figure~\ref{meta_fehm10_d25_ang_mom_groups}
shows where these stars fall on the angular momentum and cylindrical
velocity plot. The H99 streams are clumped in angular momentum and
cylindrical velocity space, but not in Galactic coordinates. The first
group of outliers, on the retrograde (left) side of the angular
momentum plot, might also be tidal debris. It has extent and isolation
similar to the H99 streams. It also forms a kinematically distinct
group in the cylindrical velocity plots (at $\avevphi \sim -300$ km
s$^{-1}$) as well as being completely made up of low metallicity stars
(see Table~\ref{meta_out1_table}; \citet{venn04} have already noted
the chemical homogeneity of the extreme retrograde stars in the solar
neighborhood). It is not, however, as tightly clumped in velocity
space as the H99 streams. The second group of outliers is located on
the prograde (lower right) side of the angular momentum plot and has
kinematics similar to disk stars, but with higher J$_z$ ($\avevphi
\sim 300$ km s$^{-1}$, $\avevR \sim 0$ km s$^{-1}$, $\avevz \sim 0$ km
s$^{-1}$).  The prograde outlier group is also very metal poor (see
Table~\ref{meta_out2_table}). Because of the closeness to the disk and
thick disk region of the angular momentum plot, it is possible that
this group is related to the disk or was accreted into the disk by
dynamical friction as in~\citet{abadi03}.

We can estimate the probability of stars from a smooth halo populating
the above regions in angular momentum space using the smooth halo
model described in Section~\ref{model}. We randomly selected 231 stars
within 2.5~kpc of the Sun from this model and counted the number of
stars in each region of the angular momentum plot where we see
outliers in the data. The size of the angular momentum region was
chosen to be large enough to enclose all the outliers and their
errors. We did not include AS Cnc in the H99 streams region because we
are not sure if it is really part of this group. (See
Section~\ref{h99_stream_prop}.) We also randomly generated errors for
the points from the model based on the distribution of errors in the
combined data set. The errors were modeled as normal distributions
with mean zero, $\sigma(J_z)= 186 \ \rm{kpc \ km \ s^{-1}}$, and
$\sigma(J_\perp) = 107 \ \rm{kpc \ km \ s^{-1}}$. This process was
repeated for 100,000 trials. Table~\ref{box_prob} gives the box
parameters, the probability that there would be as many or more stars
as there are possible stream stars in the box if the halo was entirely
smooth, and the average number of stars in the each region for a
smooth halo.

For the H99 stream, we found that there was a very low probability of
having 11 smooth halo stars in that region of the angular momentum
diagram. This result supports the conclusions of H99. In the
retrograde outlier region, there is a 7\% chance of having six or more
stars in that region of the angular momentum diagram. This set of
outliers is possibly another stream.  The simulations tell a different
story for the prograde outliers. The probability for having at least
three stars in this region of the angular momentum diagram is 70\%. In
addition, changing the criterion used to exclude thick disk stars
could eliminate the prograde group or fill in the region between the
halo distribution and the prograde group. The prograde outliers are
likely smooth halo stars (perhaps with some contribution from the
thick disk as well) rather than part of a stream.

\subsubsection{H99 Stream Properties} \label{h99_stream_prop}

The H99 streams make up 5\% of our total sample. For a subsample of
the combined data set with the same parameters as the H99 sample
($\feh \leq -1.0$ and distances less than 1.0 kpc), the fraction is
9\%. We detect 11 of 12 stars that H99 detected and possibly add one
more star to the streams (AS Cnc). The missing H99 star (HD 214925) we
group with the retrograde outlier group rather than the H99
streams. In Figure~\ref{meta_fehm10_d25_ang_mom_groups}, AS Cnc is
located very far from the rest of the H99 group in angular momentum
space (at $J_z \sim 4000$~kpc~km~s$^{-1}$ and $J_\perp \sim
3600$~kpc~km~s$^{-1}$) , but agrees with the rest of the stars in
velocity space. Therefore, we cannot say for certain whether AS Cnc is
part of the H99 group without better data on its distance and space
velocities. We include AS Cnc in our list of members of the H99
streams for completeness, not because we are certain it is a member.

\citet{katie} and \citet{2000AJ....120.1841F,fulbright2002} gave
high-dispersion abundance analyses for three of the H99 stream stars
in Table 4: HD 128279, BD+30 2611 and HD 175305. Their [Mg/Fe] and
[$\alpha$/Fe] abundances are similar to other halo stars in the solar
neighborhood, suggesting that the gas that formed these stars was
enriched mostly by Type II supernovae.

\citet{fiorentin05} has performed a similar analysis to ours using
only the B00 data and finds seven additional stars in the H99
streams. Two of the \citet{fiorentin05} stars (HD 214161 and BPS CS
22189-0007) we excluded from our sample because they were classified
by B00 as giants and had $(B-V)_0$ less than 0.9; they are likely to
have underestimated distances (see
Section~\ref{combined_rg}). \citet{fiorentin05} also include RZ Cep in
their list of members of the H99 streams. RZ Cep is a type c RR Lyrae
and thus not included in the M07 sample of RR Lyrae stars; it could be
an additional member of the stream.  Finally, five of the stars that
\citet{fiorentin05} detect (BPS CS 22948-0093, BPS CS 30339-0037, BPS
CS 29513-0031, BPS CS 29504-0044, and BPS CS 22876-0040) are
classified by B00 as turnoff stars. We did not include these stars in
our sample because the distance estimates used in B00 for turnoff
stars were based on UBV-photometry and thus not able to deal with the
evolution of turnoff stars up to the subgiant branch. This ambiguity
introduces additional uncertainty into the distance estimates for
these stars \citep[][especially their Fig. 8]{schuster04}.

We tested the $v_z$ velocity data for normality using the Shapiro-Wilk
test. The results of these tests are given in
Table~\ref{meta_fehm10_d25_vz_results}. When the H99 stream stars are
excluded from the sample, it tests positive for normality. Comparing
our results to the simulations in Section~\ref{modeling}, it is not
surprising that we were able to detect the H99 streams in this sample
using only $v_z$ velocities. We did not test the radial velocity
distribution of this sample for normality because there are not enough
stars in the sample for this test to detect structure.

\begin{figure}
\centering
\includegraphics[scale=1.0]{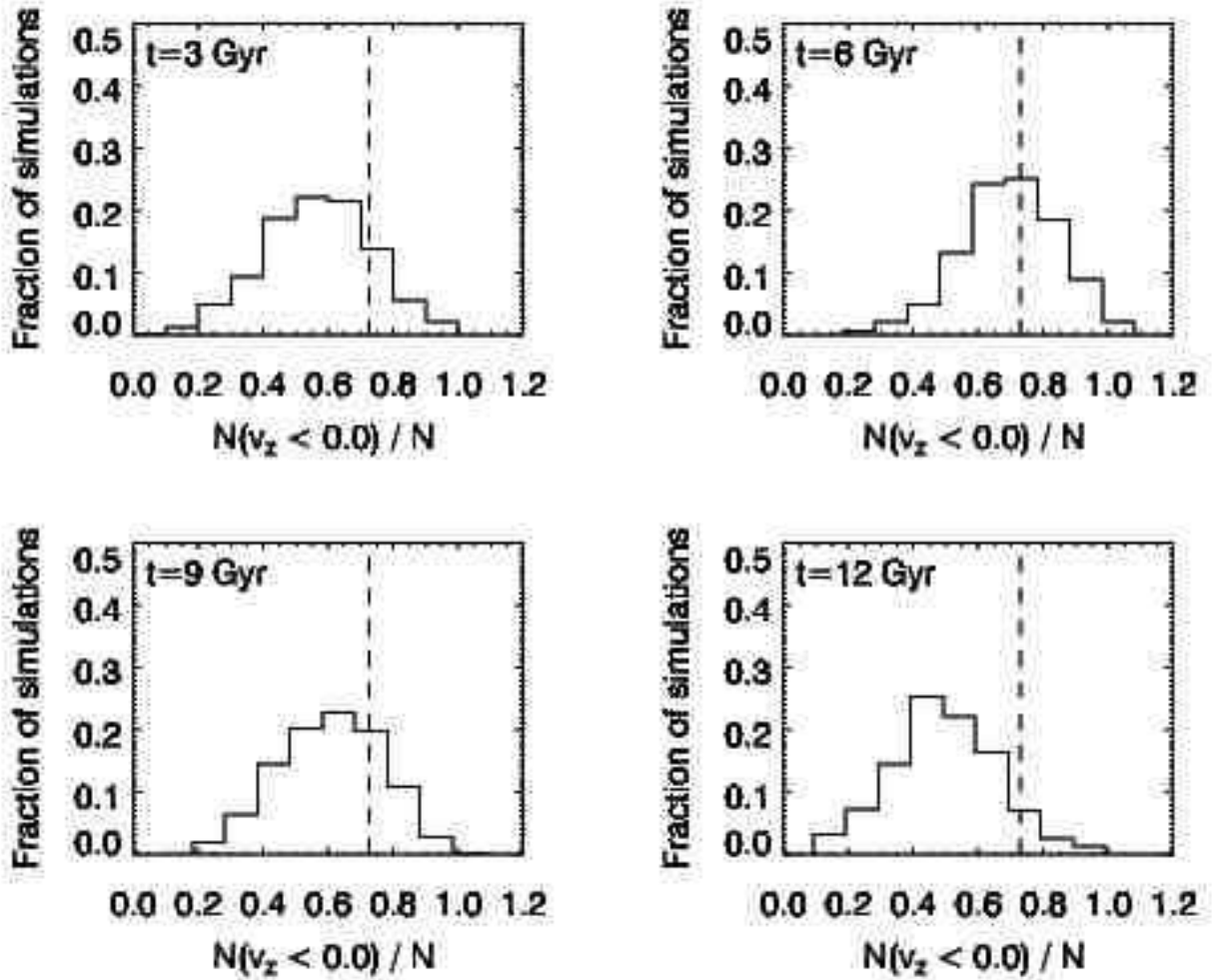}
\caption{Probability that the stream with negative $v_z$ will make
   up a given fraction of the total number of stars from the accreted
   satellite near the Sun calculated using the H99 model at four
   different times: 3, 6, 9, and 12 Gyr. The dashed line
   indicates the observed fraction of stars in this stream ($8/11 =
   0.72$).}
\label{model_vz_evolution}
\end{figure}

We can use the observed asymmetry in the $v_z$ velocity distribution
of the H99 streams along with our model of them to estimate the how
long ago the progenitor was accreted. Initially, all stars are bound,
and as time goes by, the stars are released leading to the formation
of streams.  These streams will phase-mix, and after many gigayears,
this mixing will have progressed so that half the stars will cross the
Galactic disk in each direction. In
Figure~\ref{meta_fehm10_d25_ang_mom_groups}, we see that the observed
fraction of stars in the stream with negative $v_z$ is 0.72 (8/11).
To estimate the age of the H99 streams, we selected 11 model stars
within 2.5 kpc of the Sun from our simulation and determined the
fraction in the stream with negative $v_z$. We repeated this 1000
times to build up a probability distribution. We then compared the
observed fraction of stars ($8/11=0.72$) to this probability
distribution to determine an age (Fig.~\ref{model_vz_evolution}). We
find that the observed fraction is matched by the mean simulated
fraction for a progenitor that was accreted between 6 and 9~Gyr
ago. Although we cannot determine a definite age, we can rule out the
possibility that the stream is either very young (3~Gyr) or very old
(12~Gyr). Future observations can improve this situation. By
increasing the observed total number of stars in the streams, the
width of the peak of the probability distribution in
Figure~\ref{model_vz_evolution} decreases allowing a more accurate
determination of the age of the accretion event.

\subsubsection{Retrograde Outlier Properties}

The retrograde outliers do not add significant structure to the $v_z$
velocity distribution (the sample passes the Shapiro-Wilk test if the
H99 streams, but not the retrograde outliers, are removed), but they
do have significant structure in the $v_\phi$ direction, as can be
seen in Figure~\ref{meta_fehm10_d25_ang_mom_groups}. To see if the
Shapiro-Wilk test is able to pick out these deviations from normality,
we ran various samples of $v_\phi$ data through the Shapiro-Wilk
test. The results are shown in
Table~\ref{meta_fehm10_d25_vphi_results}. Excluding the retrograde
group of stars (Fig. ~\ref{meta_fehm10_d25_ang_mom_groups}, {\em
inverted triangles}) causes the sample to pass the Shapiro-Wilk test
(test positive for normality). Indeed, all the samples that pass the
Shapiro-Wilk test exclude this group of stars, which are very far from
the rest of the $v_\phi$ velocities. Samples excluding only the
prograde outliers do not test positive for normality. Evidently
removing the prograde outliers from the sample just removes stars from
the tail of the normal distribution of the halo and does not remove
substructure.

We note that this group of stars has an extremely low value of
$J_z$. The 38 globular clusters with proper motion estimates
summarized by \citet{dinescu}, most of which are within 10 kpc of the
Galactic center, have $J_z$ values ranging from --664 to 2307
kpc~$\kms$, while the mean value of $J_z$ for our retrograde group is
--2500 kpc~$\kms$.  However, larger samples of nearby proper-motion
selected stars have produced stars with even more extreme negative
velocities, for example Kapteyn's star group, which has a mean $J_z$
around --4000 kpc~$\kms$ \citep{eggen96}.

Three of the stars from the retrograde substructure also have
high-dispersion abundance information from \citet{katie},
\citet{1995AJ....109.2757M}, and \citet{1988A&A...204..193G}: HD 6755,
HD 200654 and CD--24 1782. Interestingly, these stars show a range of
[$\alpha$/Fe]: one has roughly normal [$\alpha$/Fe] for the local halo
while the other two have low [$\alpha$/Fe] for their [Fe/H]. Note that
our retrograde group corresponds roughly to the ``extreme retrograde''
class of \citet{venn04}, and we find a similar behavior (somewhat
lower [$\alpha$/Fe] than normal) with our improved kinematical
measures.

\subsubsection{Discussion}

Including both the H99 streams and the retrograde outliers gives a
substructure fraction of 7\% for the halo.  The fraction of stream
stars in the local halo is therefore at least 5\% and possibly as high
as 7\% in this sample. \citet{gould03} produced an estimate of the
overall amount of substructure in the local halo (its ``granularity'')
using an updated version on the NLTT proper motion sample. This method
uses the fact that dominant streams in the solar neighborhood will not
show spatial substructure over such a small volume, but will produce
correlations between different components of velocity (non-zero cross
terms in the velocity dispersion tensor). These will be reflected in
the proper motion distribution. The fraction of stream stars in our
sample is consistent with the upper limit set in \citet{gould03}, that
one stream can comprise no more than 5\% of the halo.

Chiba and Beers (2000) noted that when they increase the sample size of
halo stars to 3 times that of H99 (728 stars), they only find nine
stream stars total. There are several reasons why they may not have
found more stream stars. First, the box that they use in their
Figure~15 to identify the stream is too small. Our
Figure~\ref{ang_mom_am} shows the distribution of stars in the model
within 2.5~kpc in angular momentum and velocity space. This
distribution is much larger than the box that Chiba and Beers (2000)
selects.  They also do not remove stars with thick disk kinematics
($J_z \sim 1750$ kpc km$^{-1}$ and $J_\perp \sim 0$ kpc km$^{-1}$)
from their sample, thus decreasing the percentage of H99 stream stars.
Note, however, that the clumps in the $v_\phi-v_z$ velocity space are
still apparent in their diagram, so it is still possible to pick out
the H99 streams in their larger sample. This example illustrates the
need to look not only at a angular momentum plot, but also a velocity
plot to confirm membership in a moving group.

\section{Conclusions} \label{conclusions}

We have assembled a sample of halo stars in the solar neighborhood to
look for substructure in velocity and angular momentum space. Our
sample of 231 stars includes red giants, RR Lyrae variable stars, and
RHB stars within 2.5 kpc of the Sun with [Fe/H] less than --1.0. It
was chosen to include stars with well-quantified errors and accurate
distances, space velocities, and metallicities. Understanding the
errors in the measured and derived quantities, especially distance, is
crucial for this work since they may distort any underlying
substructure.


With our data set, we confirm the existence of the streams found by
H99, which we refer to as the H99 streams. These streams have
significant structure in their velocity distribution in the $z$
direction (out of the Galactic plane). We use the results of H99 to
test how one might use $v_z$ velocity information and radial velocity
information to detect kinematic substructure in the halo. We find that
detecting the H99 streams with radial velocities alone would require a
large sample (e.g., approximately 150 stars within 2 kpc of the Sun
and within 20$^\circ$ of the Galactic poles). We also use the
structure in the velocity distribution of the H99 streams to estimate
the age of this group. From our model of the H99 progenitor, we
determine that the H99 streams' progenitor was accreted between 6 and
9~Gyr ago.

We have also discovered, in angular momentum space, two other possible
substructures, which we refer to as the retrograde and prograde
outliers. For the retrograde outliers, there is a low probability of
that region of the angular momentum diagram being occupied by six or
more smooth halo stars. Based on this evidence, the retrograde
outliers are likely members of a stream. The prograde outliers,
however, are most likely smooth halo stars (perhaps with some
contribution from the thick disk as well) rather than part of a
stream.  The retrograde outliers display significant structure in the
$v_\phi$ direction. Samples excluding the retrograde outliers pass our
test for normality in the $v_\phi$ direction. The fraction of stars in
our sample that are stream stars is between 5\% and 7\%.

For H99 streams, the [Mg/Fe] and [$\alpha$/Fe] abundances are similar
to other halo stars in the solar neighborhood, suggesting that the gas
that formed these stars was enriched mostly by Type II
supernovae. The retrograde outliers show a range of [$\alpha$/Fe]: one
has roughly normal [$\alpha$/Fe] for the local halo while the other
two have low [$\alpha$/Fe] for their [Fe/H]. Note that our retrograde
group corresponds roughly to the ``extreme retrograde'' class of
\citet{venn04}, and we find a similar behavior (somewhat lower
[$\alpha$/Fe] than normal) with our improved kinematical measures.

Although we are not the first to note that stellar debris from the
disruption of a satellite would have a double-peaked distribution in
galactocentric radial velocity \citep[see also][]{meza05}, the methods
developed in this paper add to the toolbox of kinematic methods
\citep[e.g. H99][]{h06} being developed to exploit future large
databases such as RAVE \citep{rave2006}, SDSSII/SEGUE
\citep{segue_paper}, and {\em Gaia} \citep{gaiabook} to detect kinematic
substructure in our Galaxy's halo. These tools, in conjunction with
studies of spatial over-densities in the Milky Way
\citep[e.g.][]{willman02,belokurov2006}, will provide crucial answers
to the puzzle of how our Galaxy was formed.

\acknowledgments

A. A. K. was supported by a NSF Graduate Research Fellowship during
portions of this work. H. L. M. acknowledges the support of NSF grant
AST-0098435. A. A. K. would like to thank Eric M. Wilcots for his
patience while she finished this paper, and H. L. M. thanks Bruce
Twarog and Barbara Anthony-Twarog for their helpful explanations of
the $uvby$ luminosity classifications and Kim Venn for a useful
discussion on [$\alpha$/Fe] measurements. The authors would also like
to thank the referee for his or her helpful comments.  We made
extensive use of the SIMBAD astronomical database for this project.

\appendix

\section{Transforming Local Heliocentric Coordinates to Galactocentric
Model Coordinates} \label{transform}

\begin{figure}
\centering
\includegraphics[scale=0.7]{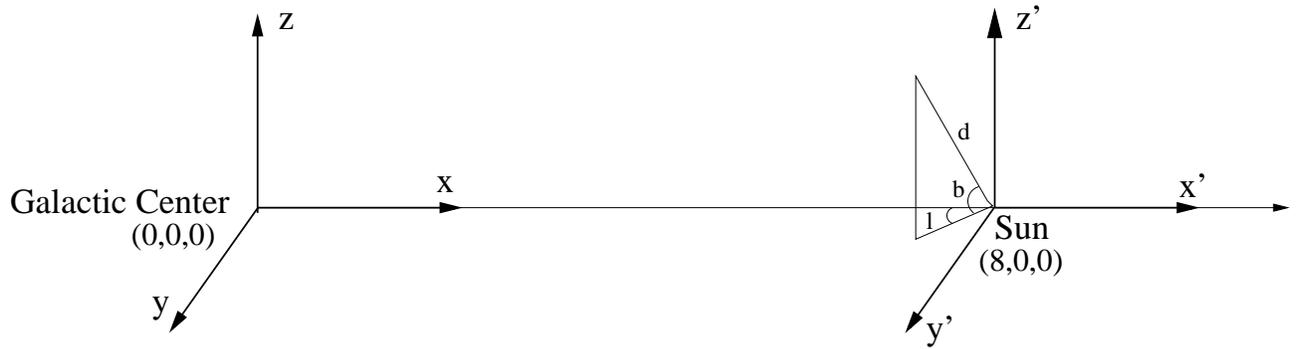}
\caption{H99 model $(x,y,z)$ and the local, Sun-centered
  $(x',y',z')$ coordinate systems. Note that both the H99 model
  coordinate system and the local, Sun-centered coordinate system are
  left-handed. The direction of Galactic rotation is in the positive
  $y'$-direction.}
\label{gal_coords}
\end{figure}

Given the wide variety of coordinate systems used in Galactic
astronomy, we explicitly derive the transformation between the local,
Sun-centered coordinate systems $(x',y',z')$ used in this paper and
the Galactocentric coordinate system used in the H99 models
$(x,y,z)$. Figure~\ref{gal_coords} illustrates the two coordinate
systems. The local velocity coordinate system is a left-handed
coordinate system with the $x'$ axis pointing away from the Galactic
center, the $y'$ axis pointing in the direction of Galactic rotation,
and the $z'$ axis pointing in the direction of the NGP. Note that the
$L_{z'}$ angular momentum vector points in the direction opposite of
the $z'$ axis. In other words, the Galaxy rotates clockwise when you
look down from the NGP. The model coordinate system used in H99 is
also left-handed.

The position of a star with respect to the Sun is given by 
\begin{eqnarray}
x' & = & - d \cos (b) \cos (l) \\
y' & = & d \cos (b) \sin (l) \\
z' & = & d \sin (b)
\end{eqnarray}
where $d$ is the distance to the star in kpc, $b$ is the Galactic latitude,
and $l$ is the Galactic longitude. The Galactocentric model
coordinates are then
\begin{eqnarray}
x & = & 8.0 \ \mathrm{kpc} + x'  \\
y & = & y' \\
z & = & z'
\end{eqnarray}
where 8.0~kpc is the distance between the Sun and the Galactic
center. 

To transform the observed space velocities to the Local
Standard of Rest (LSR) frame,
we use the equations
\begin{eqnarray}
v_{x'} & = & U + u_\odot \\
v_{y'} & = & V + v_\odot + v_{lsr} \\
v_{z'} & = & W + w_\odot 
\end{eqnarray}
where $U$, $V$, and $W$ are the space velocities directed toward the
Galactic anti-center, toward the direction of rotation, and toward the
NGP. We used the values $v_{lsr} = 220.0 \ \kms$,
$u_{\odot} = -9.0 \ \kms$, $v_{\odot} = 12.0 \ \kms$, and $w_{\odot} =
7.0 \ \kms$ \citep{blaauw,mb81} to correct for the motions of the Sun
and the LSR. These velocities are the velocities in the model frame, i.e.,
the following
\begin{eqnarray}
v_x & = & v_{x'}   \\
v_y & = & v_{y'}  \\
v_z & = & v_{z'}.
\end{eqnarray}

The angular momentum components (per unit mass) are then the
cross products
\begin{eqnarray}
J_x & = & y v_z - v_y z \\
J_y & = & z v_x - v_z x \\
J_z & = & x v_y - v_x y
\end{eqnarray}
and $J_\perp$ is $\sqrt{ J_x^2 + J_y^2}$. Note that these angular
momentum products are calculated on a left-handed system. While this
does not make a difference in $J_\perp$, a left-handed $J_z$ points in
the opposite direction as a right-handed $J_z$. In other words, the
left-handed $J_z$ of a star near the Sun is $220 \ \kms \times 8 \
\mathrm{kpc} = 1760 \ \mathrm{kpc} \ \kms$, while the traditional
right-handed $J_z$ of a star near the Sun is $-1760 \ \mathrm{kpc} \
\kms$.



\begin{deluxetable}{lrrrrrrrrrrr}
\tablewidth{0pt}
\tablecaption{Combined Sample Red Giants and RHB stars (see Section~\ref{combined_sample_sec} for details and M07 for the RR Lyrae data). \label{combined_sample_table}} 
\tablehead{ 
\colhead{Name} & 
\colhead{RV } & 
\colhead{D} & 
\colhead{$\delta$ D} &
\colhead{$\feh$} & 
\colhead{$v_{r}$} & 
\colhead{$\delta$ $v_{r}$} & 
\colhead{$v_{\phi}$} & 
\colhead{$\delta$ $v_{\phi}$} & 
\colhead{$v_{z}$} & 
\colhead{$\delta$ $v_{z}$} & 
\colhead{Source} \\
\colhead{} & 
\colhead{$\kms$ } & 
\colhead{kpc} & 
\colhead{kpc} &
\colhead{} & 
\colhead{$\kms$} &
\colhead{$\kms$} &
\colhead{$\kms$} & 
\colhead{$\kms$} & 
\colhead{$\kms$} & 
\colhead{$\kms$} &
\colhead{}}
\tablecolumns{12}
\tabletypesize{\scriptsize}
\setlength{\tabcolsep}{0.05in}
\startdata
HD 20               &  -57.4 & 0.46 & 0.07 & -1.66 & 204.40 & 14.5 & 17.21 & 14.5 & 10.89 & 3.7 & CY98\\ 
HD 97               &  76.3 & 0.47 & 0.27 & -1.38 & -259.69 & 64.2 & -67.02 & 87.0 & -124.62 & 12.2 & CY98\\ 
CD -23 72           &  20.2 & 0.55 & 0.07 & -1.12 & -7.90 & 2.5 & 163.70 & 5.4 & -24.95 & 1.3 & CY98\\ 
HD 2665             &  -378.5 & 0.24 & 0.23 & -1.87 & -170.94 & 7.2 & -115.51 & 6.4 & -30.19 & 17.6 & CY98\\ 
HD 2796             &  -60.5 & 0.72 & 0.10 & -2.35 & -109.48 & 10.3 & 75.47 & 14.8 & 34.06 & 3.3 & CY98\\ 
HD 3008             &  -80.8 & 1.54 & 0.06 & -1.87 & 53.52 & 9.0 & 101.19 & 8.8 & 59.41 & 2.3 & CY98\\ 
HD 4306             &  -67.0 & 0.56 & 0.09 & -2.72 & 148.75 & 16.3 & 168.14 & 4.7 & 84.93 & 1.7 & CY98\\ 
BD -11 145          &  -93.4 & 1.85 & 0.17 & -2.02 & -134.52 & 24.7 & 31.84 & 31.7 & 24.58 & 11.3 & CY98\\ 
HD 5426             &  27.7 & 0.72 & 0.13 & -2.33 & -42.74 & 5.2 & 26.02 & 26.2 & 2.42 & 3.2 & CY98\\ 
BD -20 170          &  -5.3 & 0.76 & 0.24 & -1.31 & 57.98 & 16.4 & 163.41 & 16.7 & 11.70 & 1.1 & CY98\\ 
CD -30 298          &  29.7 & 1.00 & 0.06 & -3.09 & 317.63 & 19.4 & 150.99 & 6.3 & -18.08 & 1.1 & CY98\\ 
HD 6446             &  62.0 & 0.62 & \nodata & -1.60 & -7.44 & \nodata & 65.70 & \nodata & 90.00 & \nodata & B00\\ 
HD 6755             &  -328.7 & 0.17 & 0.18 & -1.62 & 207.21 & 72.1 & -326.56 & 50.3 & 110.96 & 16.7 & CY98\\ 
HD 8724             &  -110.2 & 0.73 & 0.08 & -1.76 & -15.71 & 4.7 & -98.77 & 22.6 & -74.52 & 13.1 & CY98\\ 
HD 9051             &  -72.7 & 0.45 & 0.24 & -1.50 & 46.04 & 15.9 & 138.74 & 22.8 & 92.25 & 3.5 & CY98\\ 
BD -18 271          &  -209.3 & 2.38 & 0.05 & -2.06 & -164.37 & 13.6 & -56.22 & 19.6 & 167.03 & 3.8 & CY98\\ 
HD 13979            &  54.0 & 0.80 & 0.09 & -2.63 & -37.37 & 4.9 & 13.89 & 20.5 & -17.58 & 3.0 & CY98\\ 
BD -22 395          &  103.1 & 1.69 & 0.14 & -2.14 & 68.62 & 7.8 & -158.56 & 54.3 & -25.28 & 9.7 & CY98\\ 
BD -10 548          &  238.7 & 0.83 & 0.18 & -1.71 & 280.40 & 32.3 & -16.24 & 46.9 & -77.54 & 23.4 & CY98\\ 
CD -36 1052         &  306.0 & 0.72 & 0.07 & -2.19 & 87.54 & 3.9 & 137.36 & 3.7 & -267.16 & 4.8 & CY98\\ 
CD -30 1121         &  103.9 & 0.78 & 0.20 & -1.82 & 12.83 & 2.9 & 51.98 & 28.0 & -35.05 & 10.9 & CY98\\ 
HD 21022            &  110.0 & 1.19 & 0.11 & -1.99 & -25.13 & 7.7 & -44.88 & 24.2 & 8.85 & 10.3 & CY98\\ 
HD 21581            &  154.2 & 0.39 & 0.15 & -1.74 & 94.47 & 2.0 & 38.03 & 27.8 & -97.11 & 1.5 & CY98\\ 
CD -24 1782         &  118.5 & 0.73 & 0.08 & -2.37 & -137.14 & 16.8 & -276.76 & 38.1 & 8.50 & 9.0 & CY98\\ 
HD 23798            &  89.5 & 1.06 & 0.09 & -1.90 & 54.58 & 4.6 & 126.96 & 6.8 & -5.61 & 5.8 & CY98\\ 
HD 25532            &  -112.5 & 0.27 & 0.11 & -1.23 & -88.13 & 2.9 & 25.51 & 21.0 & 25.75 & 2.5 & CY98\\ 
HD 26297            &  13.5 & 0.62 & 0.08 & -1.76 & 31.91 & 3.3 & 134.86 & 8.1 & 80.72 & 7.1 & CY98\\ 
BD +06 648          &  -141.4 & 1.25 & 0.06 & -2.04 & -163.11 & 3.6 & -44.47 & 17.0 & 72.42 & 4.8 & CY98\\ 
HD 27928            &  15.1 & 0.64 & 0.16 & -2.25 & -161.56 & 24.1 & 65.80 & 23.1 & 43.86 & 8.0 & CY98\\ 
HD 29574            &  23.8 & 1.17 & \nodata & -1.55 & -213.25 & \nodata & 54.53 & \nodata & -168.71 & \nodata & CY98\\ 
HD 30229            &  304.1 & 0.61 & 0.13 & -2.32 & -122.47 & 12.5 & -74.23 & 8.0 & -92.45 & 11.8 & CY98\\ 
HD 32546            &  157.0 & 0.53 & 0.18 & -1.30 & 41.00 & 26.0 & 94.33 & 11.9 & -57.00 & 9.4 & B00\\ 
HD 268957           &  172.3 & 0.57 & 0.07 & -1.63 & 316.13 & 26.2 & 50.47 & 4.9 & -102.88 & 2.7 & CY98\\ 
HD 36702            &  121.8 & 1.12 & 0.09 & -1.86 & -51.10 & 9.0 & 90.09 & 4.8 & -53.21 & 2.8 & CY98\\ 
HD 274939           &  190.0 & 0.47 & 0.23 & -1.67 & -221.86 & 56.7 & -9.70 & 18.5 & -103.69 & 7.1 & CY98\\ 
HD 37828            &  185.0 & 0.31 & \nodata & -1.43 & 77.58 & \nodata & 70.59 & \nodata & -32.00 & \nodata & B00\\ 
HD 41667            &  302.0 & 0.53 & 0.19 & -1.18 & 272.66 & 27.6 & 72.90 & 12.5 & -32.00 & 14.2 & B00\\ 
HD 44007            &  165.3 & 0.15 & 0.18 & -1.61 & 83.42 & 5.5 & 81.16 & 8.5 & 9.53 & 7.5 & CY98\\ 
HD 74462            &  -168.1 & 0.79 & 0.12 & -1.53 & -121.39 & 1.9 & -165.61 & 40.9 & 142.10 & 29.2 & CY98\\ 
HD 82590            &  215.6 & 0.53 & 0.07 & -1.85 & -186.61 & 17.8 & -120.53 & 11.0 & -34.38 & 9.7 & CY98\\ 
HD 83212            &  109.5 & 0.68 & 0.11 & -1.49 & 10.99 & 2.1 & 102.12 & 3.9 & -24.21 & 8.6 & CY98\\ 
HD 233666           &  -65.5 & 0.53 & 0.07 & -1.65 & -70.53 & 2.2 & 151.65 & 5.1 & -2.56 & 3.0 & CY98\\ 
HD 84903            &  79.0 & 0.86 & 0.04 & -2.55 & 24.29 & 3.3 & 147.25 & 5.1 & -42.67 & 3.9 & CY98\\ 
HD 237846           &  -302.8 & 0.82 & 0.10 & -2.67 & -199.16 & 3.1 & 116.48 & 4.7 & -195.83 & 3.1 & CY98\\ 
HD 85773            &  148.8 & 1.82 & 0.05 & -2.18 & 28.81 & 7.5 & 2.31 & 6.3 & -150.74 & 13.1 & CY98\\ 
HD 88609            &  -36.2 & 1.01 & 0.05 & -2.69 & -38.55 & 3.5 & 81.27 & 8.8 & 27.57 & 3.7 & CY98\\ 
BD +30 2034         &  97.0 & 2.48 & \nodata & -1.52 & 211.71 & \nodata & 53.94 & \nodata & -42.00 & \nodata & B00\\ 
CD -30 8626         &  262.0 & 0.73 & 0.24 & -1.67 & -48.20 & 10.1 & -69.76 & 14.0 & -12.91 & 31.8 & CY98\\ 
HD 93529            &  143.0 & 0.41 & 0.26 & -1.24 & -28.59 & 4.6 & 57.89 & 12.6 & -7.21 & 22.2 & CY98\\ 
BD +04 2466         &  37.6 & 0.38 & 0.16 & -1.85 & -28.08 & 3.9 & 109.01 & 16.5 & -18.67 & 9.2 & CY98\\ 
HD 99978            &  68.0 & 0.31 & \nodata & -1.12 & -74.79 & \nodata & 185.40 & \nodata & -12.00 & \nodata & B00\\ 
CPD -70 1436        &  300.0 & 0.51 & \nodata & -2.10 & 107.58 & \nodata & -149.74 & \nodata & -65.51 & \nodata & CY98\\ 
BD +22 2411         &  34.5 & 2.23 & 0.08 & -1.95 & -15.58 & 13.1 & 51.47 & 17.6 & 7.74 & 4.6 & CY98\\ 
TY Vir              &  229.0 & 0.72 & 0.07 & -1.58 & 7.05 & 3.3 & -62.91 & 12.4 & 82.06 & 8.9 & CY98\\ 
BD -01 2582         &  0.2 & 0.36 & 0.10 & -2.32 & -76.80 & 6.7 & 63.52 & 16.3 & -95.28 & 10.6 & CY98\\ 
HD 103545           &  179.7 & 1.05 & 0.09 & -2.42 & 126.10 & 12.1 & -89.00 & 23.6 & 62.02 & 10.1 & CY98\\ 
HD 104053           &  -56.0 & 1.07 & \nodata & -1.10 & 43.12 & \nodata & 95.87 & \nodata & -76.00 & \nodata & B00\\ 
BD +09 2574         &  -48.8 & 1.12 & 0.18 & -1.95 & -82.81 & 15.4 & 23.08 & 42.6 & -125.17 & 16.6 & CY98\\ 
HD 104893           &  23.6 & 1.40 & 0.06 & -1.86 & 141.05 & 11.3 & 112.42 & 7.7 & -60.81 & 6.1 & CY98\\ 
HD 105546           &  19.3 & 0.36 & 0.07 & -1.44 & 9.38 & 1.0 & 118.81 & 8.8 & 75.91 & 4.1 & CY98\\ 
HD 106373           &  96.0 & 0.43 & 0.07 & -2.48 & 163.41 & 14.9 & 84.69 & 10.8 & 67.59 & 6.6 & CY98\\ 
HD 107752           &  220.0 & 1.36 & 0.09 & -2.68 & 142.30 & 15.2 & -175.08 & 31.1 & 118.59 & 8.8 & CY98\\ 
HD 108317           &  6.5 & 0.32 & 0.08 & -2.34 & 188.70 & 16.1 & 76.32 & 12.6 & -25.42 & 3.3 & CY98\\ 
{[MFF90]} PHI 2/2 97  &  12.0 & 2.42 & \nodata & -1.10 & 50.85 & \nodata & 134.13 & \nodata & -73.00 & \nodata & B00\\ 
HD 108577           &  -111.7 & 1.04 & 0.09 & -2.50 & 209.15 & 18.1 & -105.27 & 31.1 & -191.99 & 7.6 & CY98\\ 
BD +04 2621         &  -42.0 & 1.46 & 0.08 & -2.34 & -1.75 & 9.4 & -8.70 & 21.7 & -138.08 & 8.9 & CY98\\ 
BD +30 2294         &  56.0 & 1.81 & \nodata & -1.09 & -4.91 & \nodata & 47.01 & \nodata & 71.00 & \nodata & B00\\ 
BD +26 2368         &  100.0 & 1.26 & \nodata & -1.08 & -76.95 & \nodata & 115.37 & \nodata & 110.00 & \nodata & B00\\ 
HD 109823           &  8.0 & 0.83 & \nodata & -1.68 & -11.10 & \nodata & 150.99 & \nodata & 21.00 & \nodata & B00\\ 
HD 110184           &  140.1 & 1.22 & 0.04 & -2.31 & 12.84 & 5.2 & 104.94 & 5.0 & 119.10 & 1.6 & CY98\\ 
HD 110281           &  141.6 & 2.48 & \nodata & -1.75 & -276.64 & \nodata & -101.01 & \nodata & -34.03 & \nodata & CY98\\ 
CD -27 8864         &  247.0 & 1.77 & \nodata & -1.71 & 82.62 & \nodata & -126.86 & \nodata & 37.00 & \nodata & B00\\ 
BD +33 2273         &  41.0 & 1.46 & \nodata & -1.05 & 66.17 & \nodata & 81.06 & \nodata & 63.00 & \nodata & B00\\ 
HD 112126           &  -62.0 & 1.30 & \nodata & -1.52 & 41.24 & \nodata & 17.45 & \nodata & -32.00 & \nodata & B00\\ 
BD +10 2495         &  252.9 & 0.78 & 0.17 & -2.14 & 136.83 & 35.0 & 175.60 & 4.0 & 288.68 & 7.3 & CY98\\ 
BD +12 2547         &  5.7 & 1.47 & 0.11 & -2.07 & -233.94 & 24.4 & -66.87 & 29.9 & -91.91 & 10.9 & CY98\\ 
HD 115444           &  -27.6 & 0.78 & 0.08 & -2.63 & -156.34 & 12.2 & 63.50 & 13.6 & 13.33 & 5.7 & CY98\\ 
BD +03 2782         &  32.0 & 1.82 & \nodata & -2.02 & 146.92 & \nodata & 38.17 & \nodata & 57.00 & \nodata & B00\\ 
HD 118055           &  -101.0 & 1.13 & 0.06 & -1.76 & 100.29 & 5.6 & 161.94 & 8.3 & -98.11 & 3.3 & CY98\\ 
BD +18 2757         &  -22.2 & 1.34 & 0.09 & -2.52 & -28.41 & 5.5 & 29.14 & 21.5 & -25.34 & 1.8 & CY98\\ 
HD 119516           &  -287.0 & 0.51 & 0.07 & -2.49 & 145.36 & 5.4 & 143.15 & 7.5 & -250.47 & 1.6 & CY98\\ 
HD 121135           &  126.4 & 0.87 & 0.19 & -1.83 & -2.38 & 13.2 & 62.53 & 28.9 & 118.71 & 1.7 & CY98\\ 
HD 121261           &  99.4 & 1.30 & 0.14 & -1.52 & 78.77 & 19.7 & 5.80 & 21.9 & 42.21 & 4.2 & CY98\\ 
HD 122563           &  -24.8 & 0.31 & 0.06 & -2.55 & 142.80 & 8.3 & -20.03 & 14.8 & 27.63 & 2.7 & CY98\\ 
HD 122956           &  166.3 & 0.36 & 0.11 & -1.74 & -21.75 & 9.4 & 19.50 & 16.4 & 122.18 & 1.2 & CY98\\ 
HD 124358           &  325.0 & 1.13 & 0.12 & -1.98 & 104.03 & 38.7 & -290.72 & 55.5 & 313.68 & 8.8 & CY98\\ 
BD +09 2860         &  -19.0 & 1.04 & 0.07 & -1.67 & -161.42 & 13.1 & 105.27 & 12.5 & -97.97 & 7.1 & CY98\\ 
BD +09 2870         &  -120.7 & 1.35 & 0.06 & -2.39 & 306.41 & 16.8 & 22.15 & 15.2 & 22.55 & 8.0 & CY98\\ 
BD +01 2916         &  -12.7 & 1.92 & 0.07 & -1.61 & 81.75 & 8.7 & 3.16 & 18.1 & 13.73 & 4.7 & CY98\\ 
BD +08 2856         &  64.5 & 2.45 & 0.07 & -2.02 & -261.61 & 19.7 & 116.59 & 15.6 & -62.71 & 11.2 & CY98\\ 
HD 126238           &  247.0 & 0.24 & 0.14 & -1.85 & -209.36 & 4.3 & 42.51 & 5.6 & -39.14 & 16.8 & CY98\\ 
HD 126587           &  149.2 & 0.62 & 0.08 & -2.79 & -121.34 & 2.2 & 37.26 & 10.8 & -3.52 & 8.2 & CY98\\ 
HD 128279           &  -81.7 & 0.16 & 0.22 & -2.22 & -21.13 & 15.4 & 140.32 & 27.3 & -252.33 & 49.9 & CY98\\ 
BD +30 2611         &  -278.2 & 1.11 & 0.10 & -1.26 & 13.91 & 9.0 & 155.60 & 4.6 & -276.18 & 4.7 & CY98\\ 
BD +18 2976         &  -173.1 & 1.69 & 0.07 & -2.42 & 24.24 & 10.7 & -155.61 & 26.5 & -77.57 & 7.6 & CY98\\ 
HD 135148           &  -91.9 & 1.11 & 0.05 & -1.88 & 135.17 & 6.5 & 127.27 & 6.8 & 13.10 & 5.2 & CY98\\ 
HD 135449           &  -23.0 & 0.45 & 0.07 & -1.47 & 155.07 & 14.4 & -0.11 & 18.3 & 76.97 & 7.5 & CY98\\ 
HD 136316           &  -45.0 & 0.50 & 0.08 & -2.00 & 96.81 & 11.9 & 147.78 & 11.4 & -87.96 & 7.5 & CY98\\ 
BD +01 3070         &  -329.4 & 0.50 & 0.23 & -1.85 & 343.77 & 27.3 & 273.99 & 12.6 & -112.58 & 26.0 & CY98\\ 
HD 141531           &  2.6 & 1.29 & 0.13 & -1.57 & -182.48 & 23.1 & -66.32 & 36.9 & -52.25 & 10.7 & CY98\\ 
BD +05 3098         &  -160.5 & 1.23 & 0.12 & -2.40 & 28.43 & 11.1 & -205.59 & 47.2 & -65.18 & 10.9 & CY98\\ 
BD +11 2998         &  50.2 & 0.42 & 0.07 & -1.22 & -54.74 & 1.5 & 36.03 & 15.3 & 155.52 & 8.6 & CY98\\ 
BD +09 3223         &  67.0 & 0.50 & 0.07 & -2.41 & -191.27 & 9.6 & 32.33 & 16.3 & 9.54 & 3.4 & CY98\\ 
BD +17 3248         &  -145.6 & 0.51 & 0.21 & -2.07 & 59.78 & 7.4 & 50.73 & 20.2 & 16.20 & 15.3 & CY98\\ 
CD -68 1881         &  134.0 & 1.19 & 0.33 & -1.81 & -62.26 & 28.0 & 153.98 & 19.6 & -173.00 & 50.7 & B00\\ 
HD 165195           &  -0.2 & 0.65 & 0.03 & -2.42 & -151.91 & 6.1 & 14.94 & 7.2 & -30.55 & 2.3 & CY98\\ 
HD 166161           &  68.4 & 0.20 & 0.15 & -1.27 & -146.25 & 9.9 & 72.32 & 26.6 & 9.01 & 1.2 & CY98\\ 
HD 175305           &  -181.0 & 0.12 & 0.19 & -1.54 & 32.51 & 17.0 & 128.72 & 10.7 & -222.72 & 30.1 & CY98\\ 
HD 174578           &  -4.0 & 1.10 & 0.36 & -1.69 & -1.42 & 10.0 & 350.17 & 37.9 & 94.00 & 28.3 & B00\\ 
HD 184266           &  -348.4 & 0.22 & 0.07 & -1.87 & 298.37 & 1.0 & -50.17 & 10.4 & -98.31 & 14.5 & CY98\\ 
HD 184711           &  101.6 & 1.01 & 0.04 & -2.30 & -74.43 & 2.4 & -0.15 & 9.3 & -104.23 & 4.4 & CY98\\ 
HD 232078           &  -391.0 & 0.83 & \nodata & -1.61 & 137.49 & \nodata & -136.65 & \nodata & -75.00 & \nodata & B00\\ 
HD 186478           &  30.9 & 1.02 & 0.06 & -2.45 & -192.92 & 9.2 & -137.86 & 25.3 & -67.16 & 6.0 & CY98\\ 
HD 187111           &  -181.0 & 0.61 & 0.18 & -1.95 & 137.06 & 2.9 & 24.79 & 21.5 & -99.15 & 28.6 & CY98\\ 
BD -18 5550         &  -126.2 & 0.74 & 0.08 & -2.84 & 36.96 & 5.1 & -98.39 & 22.6 & -96.32 & 12.6 & CY98\\ 
HD 190287           &  135.0 & 0.14 & 0.23 & -1.09 & -136.43 & 3.5 & 129.59 & 26.3 & -63.15 & 1.9 & CY98\\ 
HD 195636           &  -257.8 & 0.59 & 0.07 & -2.82 & -36.60 & 14.4 & -141.82 & 17.0 & 166.43 & 4.5 & CY98\\ 
BD -17 6036         &  19.2 & 1.27 & 0.09 & -2.70 & -200.00 & 17.0 & 17.60 & 22.3 & 52.15 & 11.6 & CY98\\ 
BD -15 5781         &  -76.4 & 1.93 & 0.08 & -2.47 & 16.04 & 7.5 & 26.02 & 18.7 & -61.81 & 12.6 & CY98\\ 
BD -14 5890         &  117.5 & 0.93 & 0.19 & -2.01 & -267.26 & 30.9 & -19.80 & 60.3 & -130.64 & 13.5 & CY98\\ 
CD -37 14010        &  -200.0 & 1.68 & \nodata & -2.55 & 244.98 & \nodata & -94.42 & \nodata & -33.00 & \nodata & B00\\ 
HD 200654           &  -48.0 & 0.46 & 0.06 & -2.79 & 382.45 & 21.2 & -344.74 & 34.4 & -225.92 & 15.9 & CY98\\ 
BD -03 5215         &  -293.7 & 0.73 & 0.07 & -1.72 & 147.76 & 4.3 & -30.82 & 5.8 & 96.55 & 7.1 & CY98\\ 
HD 204543           &  -98.2 & 0.72 & 0.11 & -1.69 & -32.91 & 9.8 & 47.15 & 14.9 & -5.24 & 7.5 & CY98\\ 
HD 205547           &  47.0 & 1.10 & \nodata & -1.85 & -190.90 & \nodata & 46.41 & \nodata & 103.00 & \nodata & B00\\ 
HD 206739           &  -57.8 & 0.57 & \nodata & -1.58 & 78.66 & \nodata & 115.51 & \nodata & -54.51 & \nodata & CY98\\ 
BD -09 5831         &  14.5 & 1.90 & 0.13 & -1.87 & -68.55 & 13.2 & 126.24 & 17.0 & -56.01 & 9.2 & CY98\\ 
HD 235766           &  -314.0 & 0.92 & \nodata & -2.35 & -315.17 & \nodata & 6.82 & \nodata & 3.00 & \nodata & B00\\ 
HD 214362           &  -92.3 & 0.49 & 0.07 & -2.20 & 321.17 & 21.0 & -20.30 & 15.4 & -133.39 & 15.4 & CY98\\ 
HD 214925           &  -328.0 & 2.15 & 0.08 & -2.14 & 93.56 & 9.5 & -289.91 & 36.1 & 143.00 & 14.0 & B00\\ 
HD 216143           &  -115.9 & 0.69 & 0.07 & -2.20 & -327.79 & 25.1 & -29.90 & 15.9 & 76.35 & 2.4 & CY98\\ 
HD 218857           &  -169.6 & 0.41 & 0.19 & -2.15 & -129.95 & 32.0 & 44.21 & 23.7 & 153.77 & 1.2 & CY98\\ 
HD 220662           &  -78.1 & 1.87 & 0.08 & -1.59 & 0.33 & 11.2 & 10.05 & 17.2 & 37.78 & 5.0 & CY98\\ 
HD 220838           &  -22.8 & 1.43 & 0.07 & -1.72 & -24.64 & 8.4 & 149.05 & 8.0 & 22.65 & 3.1 & CY98\\ 
HD 221170           &  -121.9 & 0.69 & 0.07 & -2.01 & -151.22 & 8.6 & 93.53 & 3.3 & -62.75 & 8.8 & CY98\\ 
HD 222434           &  13.4 & 0.97 & 0.14 & -1.56 & -98.84 & 12.2 & 43.08 & 26.1 & 10.87 & 3.6 & CY98\\ 
\enddata
\end{deluxetable}

\clearpage

\begin{deluxetable}{crr}
\tablewidth{0pt}
\tablecaption{Smooth Halo Model Parameters \label{smo_params}}
\tablehead{ 
  \colhead{} &
  \colhead{With H99 Streams} &
  \colhead{Without H99 Streams} }
\tablecolumns{3}
\startdata
$\avevR$	&  $ -0.93 \pm 10.20$	& $ -3.57 \pm 10.63$ \\
$\sigma_R$	&  $154.96 \pm 7.21$	& $157.30 \pm 7.52$ \\
$\avevphi$	&  $ 28.62 \pm 7.19$	& $23.42 \pm 7.41$ \\
$\sigma_\phi$	&  $109.33 \pm 5.09$	& $109.64 \pm 5.24$ \\
$\avevz$	&  $ -6.43  \pm 6.66$	&   $-1.32 \pm 5.66$ \\     
$\sigma_z$	&  $101.23  \pm 4.71$	&  $83.75 \pm 4.00$ \\
\enddata
\end{deluxetable}


\begin{deluxetable}{lll}
\tablewidth{0pt}
\tablecaption{Shapiro-Wilk Results for H99 data \label{sw_h99} }
\tablehead{ \colhead{Data Set} & $N_{stars}$ & \colhead{p-value} }
\startdata
All Stars			&  101 &  0.009 \\
Excluding H99 Streams		&  94  & 0.05 \\
Excluding H99 Streams \& HD 124358	& 93   & 0.19    \\
\enddata
\end{deluxetable}

\clearpage

\begin{deluxetable}{lrrrrrrrrrrr}
\tablewidth{0pt}
\tablecaption{H99 Stream Stars in Combined Data Set \label{meta_strm_table}}
\tablehead{ 
\colhead{Name} & 
\colhead{RV} & 
\colhead{D} & 
\colhead{$\delta$ D} &
\colhead{$\feh$}  & 
\colhead{$v_r$} &
\colhead{$\delta v_r$} &
\colhead{$v_\phi$} & 
\colhead{$\delta v_\phi$} &
\colhead{$v_z$}  &
\colhead{$\delta v_z$}  &
\colhead{Source} \\	
\colhead{} & 
\colhead{km s$^{-1}$} & 
\colhead{kpc}  & 
\colhead{kpc}  & 
\colhead{} &  
\colhead{km s$^{-1}$} & 
\colhead{km s$^{-1}$} & 
\colhead{km s$^{-1}$} & 
\colhead{km s$^{-1}$} & 
\colhead{km s$^{-1}$} & 
\colhead{km s$^{-1}$} &
\colhead{}}
\tablecolumns{12} 
\tabletypesize{\scriptsize} 
\setlength{\tabcolsep}{0.02in} 
\startdata
CD -36 1052          & 306.0 & 0.72 & 0.07 & -2.19 & 87.54 & 3.9 & 137.36 & 3.7 & -267.16 & 4.8 & CY98 \\ 
AS CNC               & 258.0 & 2.31 & 0.16 & -1.89 & 16.09 & 39.7 & 38.16 & 54.2 & 363.00 & 49.4 & M06 \\ 
TT CNC               & 47.0 & 1.22 & 0.09 & -1.58 & 106.04 & 12.0 & 87.25 & 13.8 & -231.10 & 22.0 & M06 \\ 
TT LYN               & -67.0 & 0.65 & 0.05 & -1.76 & 124.42 & 13.0 & 106.61 & 8.8 & -233.30 & 13.9 & M06 \\ 
HD 237846            & -302.8 & 0.82 & 0.10 & -2.67 & -199.16 & 3.1 & 116.48 & 4.7 & -195.83 & 3.1 & CY98 \\ 
BD +10 2495          & 252.9 & 0.78 & 0.17 & -2.14 & 136.83 & 35.0 & 175.60 & 4.0 & 288.68 & 7.3 & CY98 \\ 
HD 119516            & -287.0 & 0.51 & 0.07 & -2.49 & 145.36 & 5.4 & 143.15 & 7.5 & -250.47 & 1.6 & CY98 \\ 
HD 128279            & -81.7 & 0.16 & 0.22 & -2.22 & -21.13 & 15.4 & 140.32 & 27.3 & -252.33 & 49.9 & CY98 \\ 
BD +30 2611          & -278.2 & 1.11 & 0.10 & -1.26 & 13.91 & 9.0 & 155.60 & 4.6 & -276.18 & 4.7 & CY98 \\ 
AR SER               & 127.0 & 1.71 & 0.12 & -1.78 & 98.02 & 21.3 & 100.60 & 24.8 & 313.40 & 23.1 & M06 \\ 
HD 175305            & -181.0 & 0.12 & 0.19 & -1.54 & 32.51 & 17.0 & 128.72 & 10.7 & -222.72 & 30.1 & CY98 \\ 
XZ CYG               & -146.0 & 0.50 & 0.04 & -1.52 & 26.64 & 4.5 & 152.18 & 11.9 & -232.50 & 15.2 & M06 \\ 
\enddata
\end{deluxetable}

\begin{deluxetable}{lrrrrrrrrrrr}
\tablewidth{0pt}
\tablecaption{Retrograde Outlier Stars in Combined Data Set \label{meta_out1_table}}
\tablehead{ 
\colhead{Name} & 
\colhead{RV} & 
\colhead{D} & 
\colhead{$\delta$ D} &
\colhead{$\feh$}  & 
\colhead{$v_r$} &
\colhead{$\delta v_r$} &
\colhead{$v_\phi$} & 
\colhead{$\delta v_\phi$} &
\colhead{$v_z$}  &
\colhead{$\delta v_z$}  &
\colhead{Source} \\	
\colhead{} & 
\colhead{km s$^{-1}$} & 
\colhead{kpc}  & 
\colhead{kpc}  & 
\colhead{} &  
\colhead{km s$^{-1}$} & 
\colhead{km s$^{-1}$} & 
\colhead{km s$^{-1}$} & 
\colhead{km s$^{-1}$} & 
\colhead{km s$^{-1}$} & 
\colhead{km s$^{-1}$} &
\colhead{}}
\tablecolumns{12} 
\tabletypesize{\scriptsize} 
\setlength{\tabcolsep}{0.02in} 
\startdata
HD 6755              & -328.7 & 0.17 & 0.18 & -1.62 & 207.21 & 72.1 & -326.56 & 50.3 & 110.96 & 16.7 & CY98 \\ 
CD -24 1782          & 118.5 & 0.73 & 0.08 & -2.37 & -137.14 & 16.8 & -276.76 & 38.1 & 8.50 & 9.0 & CY98 \\ 
HD 124358            & 325.0 & 1.13 & 0.12 & -1.98 & 104.03 & 38.7 & -290.72 & 55.5 & 313.68 & 8.8 & CY98 \\ 
RV CAP               & -84.0 & 1.06 & 0.07 & -1.72 & -69.03 & 9.5 & -279.92 & 34.3 & -193.70 & 18.3 & M06 \\ 
HD 200654            & -48.0 & 0.46 & 0.06 & -2.79 & 382.45 & 21.2 & -344.74 & 34.4 & -225.92 & 15.9 & CY98 \\ 
HD 214925            & -328.0 & 2.15 & 0.08 & -2.14 & 93.56 & 9.5 & -289.91 & 36.1 & 143.00 & 14.0 & B00 \\ 
\enddata
\end{deluxetable}

\begin{deluxetable}{lrrrrrrrrrrr}
\tablewidth{0pt}
\tablecaption{Prograde Outlier Stars in Combined Data Set \label{meta_out2_table}}
\tablehead{ 
\colhead{Name} & 
\colhead{RV} & 
\colhead{D} & 
\colhead{$\delta$ D} &
\colhead{$\feh$}  & 
\colhead{$v_r$} &
\colhead{$\delta v_r$} &
\colhead{$v_\phi$} & 
\colhead{$\delta v_\phi$} &
\colhead{$v_z$}  &
\colhead{$\delta v_z$}  &
\colhead{Source} \\	
\colhead{} & 
\colhead{km s$^{-1}$} & 
\colhead{kpc}  & 
\colhead{kpc}  & 
\colhead{} &  
\colhead{km s$^{-1}$} & 
\colhead{km s$^{-1}$} & 
\colhead{km s$^{-1}$} & 
\colhead{km s$^{-1}$} & 
\colhead{km s$^{-1}$} & 
\colhead{km s$^{-1}$} &
\colhead{}}
\tablecolumns{12} 
\tabletypesize{\scriptsize} 
\setlength{\tabcolsep}{0.02in} 
\startdata
TV LEO               & -97.0 & 1.84 & 0.13 & -1.97 & -113.72 & 30.4 & 326.94 & 24.2 & -22.90 & 20.9 & M06 \\ 
BD +01 3070          & -329.4 & 0.50 & 0.23 & -1.85 & 343.77 & 27.3 & 273.99 & 12.6 & -112.58 & 26.0 & CY98 \\ 
HD 174578            & -4.0 & 1.10 & 0.36 & -1.69 & -1.42 & 10.0 & 350.17 & 37.9 & 94.00 & 28.3 & B00 \\ 
\enddata
\end{deluxetable}

\clearpage

\begin{deluxetable}{lclllr}
\tablewidth{0pt}
\tablecaption{Probability of Finding Halo Stars in Various Regions of the Angular Momentum Diagram \label{box_prob}} 
\tablehead{
\colhead{} &
\colhead{} &
\colhead{Lower Left } &
\colhead{Upper Right} &
\colhead{} &
\colhead{} \\
\colhead{Group} &
\colhead{Number of Stars} &
\colhead{Corner} &
\colhead{Corner} &
\colhead{$P(n \ge n_*$)} &
\colhead{$<n>$} }
\tablecolumns{6}
\startdata
H99 Streams	     & 11	& $[500,1400]$	& $[1500,2500]$   & 0.00024   &  2.81 \\
Retrograde Outliers  & 6	& $[-3000,0]$	& $[-1750,2500]$  & 0.069     &  2.86  \\
Prograde Outliers    & 3	& $[2000,30]$   & $[3000,1200]$   & 0.69      &  3.55  \\
\enddata
\end{deluxetable}

\begin{deluxetable}{rccccr}
\tablewidth{0pt}
\tablecaption{Shapiro-Wilk test results for combined sample $v_z$ data ($D \leq 2.5 \ \mathrm{kpc}$ and $\feh \leq -1.0$) \label{meta_fehm10_d25_vz_results}}
\tablehead{ 
\colhead{N} & \multicolumn{3}{c}{Excluded Groups} & \colhead{p-value}  \\ 
 \colhead{} & \colhead{Prograde} & \colhead{Retrograde}  & \colhead{H99 streams} & \colhead{} \\
 \colhead{} & \colhead{$\blacktriangle$} & \colhead{$\blacktriangledown$} & \colhead{$\bullet$} & \colhead{} }
\tablecolumns{5}
\startdata
231 & \nodata & \nodata & \nodata & 0.002\\ 
219 & \nodata & \nodata & X & 0.584\\ 
216 & X & \nodata & X & 0.584\\ 
210 & X & X & X & 0.877\\ 
213 & \nodata & X & X & 0.850\\ 
\enddata
\end{deluxetable}

\begin{deluxetable}{rccccr}
\tablewidth{0pt}
\tablecaption{Shapiro-Wilk test results for combined sample $v_\phi$ data ($D \leq 2.5 \ \mathrm{kpc}$ and $\feh \leq -1.0$) \label{meta_fehm10_d25_vphi_results}}
\tablehead{ 
\colhead{N} & \multicolumn{3}{c}{Excluded Groups} & \colhead{p-value}  \\ 
 \colhead{} & \colhead{Prograde} & \colhead{Retrograde}  & \colhead{H99 streams} & \colhead{} \\
 \colhead{} & \colhead{$\blacktriangle$} & \colhead{$\blacktriangledown$} & \colhead{$\bullet$} & \colhead{} }
\tablecolumns{5}
\startdata
231 & \nodata & \nodata & \nodata & 0.00014\\ 
228 & X & \nodata & \nodata & 1.9e-06\\ 
222 & X & X & \nodata & 0.01\\ 
225 & \nodata & X & \nodata & 0.21\\ 
219 & \nodata & \nodata & X & 0.00061\\ 
216 & X & \nodata & X & 1.1e-05\\ 
210 & X & X & X & 0.068\\ 
213 & \nodata & X & X & 0.41\\ 
\enddata
\end{deluxetable}

\clearpage



\end{document}